\documentclass{amsart}

\usepackage[left=2.50cm, right=2.50cm, top=2.50cm, bottom=2.50cm]{geometry}

\usepackage{amsmath}
\usepackage{amsfonts}
\usepackage{amssymb}
\usepackage{graphicx}

\usepackage[mediumspace,mediumqspace,squaren]{SIunits}

\newcommand{\bs}{\boldsymbol}
\newcommand{\mb}{\mathbf}

\begin{document}

\title[Multiscale rheology of wadsleyite]{Multiscale modeling of the effective 
	viscoplastic behavior of $\text{Mg}_2\text{SiO}_4$  wadsleyite: bridging atomic 
	and polycrystal scales}

\author[O. Castelnau et al.]{O. Castelnau $^{1*}$, K. Derrien $^1$, S. Ritterbex $^{2,3}$, P. Carrez $^2$, P. Cordier $^{2,4}$, H. Moulinec $^5$}

\address{$^1$ Laboratoire PIMM, Arts et Metiers Institute of Technology, CNRS, Cnam, HESAM University, 151 boulevard de l’Hopital, 75013 Paris (France)}
\address{$^*$ Corresponding author}
\email{olivier.castelnau@ensam.eu}

\address{$^2$ Univ. Lille, CNRS, INRAE, Centrale Lille, UMR 8207 - UMET - Unité Matériaux et Transformations, F-59000 Lille, France}

\address{$^3$ Geodynamics Research Center, Ehime University, 2-5 Bunkyo-cho, Matsuyama 790-8577, Japan}

\address{$^4$ Institut Universitaire de France, 1 rue Descartes, 75005 Paris, France}

\address{$^5$ Aix Marseille Univ, CNRS, Centrale Marseille, LMA UMR 7031, Marseille, France}

\thanks{This work was supported by the European Research Council under the Seventh Framework Programme (FP 7), ERC [grant number 290424 RheoMan] and under the Horizon 2020 research and innovation programme [grant number 787198 TimeMan].}


\subjclass{XXXXXXXX}


\begin{abstract} 
	The viscoplastic behavior of polycrystalline  Mg$_{2}$SiO$_{4}$ wadsleyite 
	aggregates, a major high pressure phase of the mantle transition zone of the 
	Earth (depth range: $410-520\;\kilo\meter$), is obtained by properly bridging 
	several scale transition models. 
	At the very fine nanometric scale corresponding to the dislocation core 
	structure, the behavior of thermally activated plastic slip is modeled 
	for strain-rates relevant for laboratory experimental conditions, at high 
	pressure and for a wide range of temperatures, based on the 
	Peierls-Nabarro-Galerkin model. 
	Corresponding single slip reference resolved shear stresses and associated 
	constitutive equations are deduced from Orowan's equation in order to describe 
	the average viscoplastic behavior at the grain scale, for the easiest slip 
	systems. 
	These data have been implemented in two grain-polycrystal scale transition 
	models, a mean-field one (the recent Fully-Optimized Second-Order Viscoplastic 
	Self-Consistent scheme of \cite{Song2018}) allowing rapid evaluation of the 
	effective viscosity of polycrystalline aggregates, and a full-field  (FFT based 
	\cite{Suquet201264} \cite{Moulinec199869}) method allowing investigating stress 
	and strain-rate localization in typical microstructures and heterogeneous 
	activation of slip systems within grains. 
	Calculations have been performed at pressure and temperatures relevant for 
	in-situ conditions. 
	Results are in very good agreement with available mechanical tests conducted at 
	strain-rates typical for laboratory experiments.
\end{abstract}

\maketitle


\section{Introduction}

The flow of rocks in the Earth’s mantle controls many large-scale geodynamic processes. 
Among them, mantle convection is of primary importance since it constitutes the 
main mechanism for the Earth to evacuate internal heat, and it drives 
continental drift and associated seismic events. 
Quantitative modeling of mantle convection, which also allows investigation of 
past dynamics and prediction of future events, depends on our understanding of 
the rheology of rocks under thermo-mechanical conditions that prevail in the 
Earth’s interior. 

The Earth's mantle extends to ca. $2900\;\kilo\meter$ depth where pressure 
reaches $130\;\giga\pascal$ and temperature is exceeding $3000\kelvin$. 
In response to increasing pressure and temperature with depth, minerals observed in rocks from the upper mantle transform at depth to denser assemblages. 
These phase transitions are responsible to global discontinuities of the 
velocities of seismic waves across the corresponding depths. 
The $410\;\kilo\meter$ discontinuity is generally thought to be caused by the 
phase transition of olivine (the low pressure phase of (Mg, Fe)$_2$SiO$_4$) to 
wadsleyite and the $520\;\kilo\meter$ discontinuity by the phase transition of 
wadsleyite to ringwoodite. 
These phase transitions give rise to the so-called transition zone which ends 
at the $670\;\kilo\meter$ discontinuity where the lower mantle begins.

The transition zone is a major layer between the upper and lower mantles and is 
expected to influence the whole mantle convection depending on its 
rheological properties.
Information on the mechanical properties of rocks come primarily from 
laboratory mechanical tests. 
This experimental approach is however rendered very challenging since the $P$, 
$T$ conditions are in the range of $15\;\giga\pascal$ and $1600\kelvin$ for 
wadsleyite in the upper transition zone. 
Furthermore, rheological laws deduced from laboratory experiments at 
strain-rates of typically $10^{-5}\;\second^{-1}$ need to be extrapolated by 
$\sim10$ orders of magnitudes to the extremely low strain-rate conditions (ca. 
$10^{-15}\;\second^{-1}$) prevailing during Earth’s mantle convection.

In that context, the computational approach is an alternative to infer the viscoplastic behavior of mantle rocks and offers the potentiality to tackle the extremely low strain-rate conditions issue, provided all relevant and physical-based deformation mechanisms at play in the mantle are properly taken into account.
However, this requires bridging several characteristic length scales, from sub-nanometer to sub-meter.
To be able to glide, dislocations must overcome their intrinsic lattice friction, which strongly depends on their structure at the atomic scale (sub-nm).
Core structures of dislocations belonging to given slip systems can be calculated using the Peierls-Nabarro-Galerkin method \cite{Denoual20071915}, relying on first principle simulations of generalized stacking fault (GSF) surfaces. 
This allows addressing accurately the effect of pressure on atomic bonding. 
Intrinsic lattice friction is then calculated and quantified by the Peierls potential. 
At finite temperatures, dislocation glide mobility results from thermally-activated motion of dislocations
over their Peierls potentials.
The obtained energy barriers for dislocation glide can then be combined with Boltzmann statistics to provide a constitutive relation at the grain level (mm scale), for each available slip systems \cite{RitterbexPePi2015,Ritterbex20162085}.

To address the rheology of polycrystalline aggregates (sub-m scale), a second scale transition, from the grain to the polycrystal, must be carried out.
In minerals, this is another difficult task as few independent slip systems are generally available.
For example, olivine exhibits less than 4 independent dislocation slip systems 
leading to an extreme viscoplastic anisotropy at the grain scale 
and, as a consequence, a quite challenging application of mean-field 
homogenization methods.
Indeed, as shown by Pierre Gilormini \cite{Gilormini-1995b,Gilormini-1995a}, many earlier 
homogenization methods provides an unrealistically stiff (i.e. violating a 
rigourous upper bound) estimation of the effective rheology.
Using the more advanced Partially Optimized Second-Order (POSO) Self-Consitent 
(SC) estimate proposed by \cite{Ponte2002a737}, it has been shown  that the 
overall viscoplastic behavior (flow stress \textit{and} stress sensitivity) in 
olivine is controlled by the behavior of accommodation mechanisms (dislocation 
climb, diffusion, grain boundary sliding, ...) which are not clearly identified 
yet  
\cite{Castelnau-et-al-2008a,Castelnau-et-al-2008b,Castelnau-et-al-2010,Detrez-et-al-JMPS-2015}.
Moreover, rocks are subjected to very large strain during mantle 
convection, inducing the development of pronounced crystallographic textures 
that can be partly characterized by the anisotropic velocities of seismic waves 
\cite{Montagner2015}.
The associated effective viscoplatic anisotropy may strongly influence the 
mantle flow pattern in-situ \cite{Blackman-etal-GJI2017,Ribe-etal-2019}.
However, the prediction of in-situ texture distribution is tedious as, besides 
temperature-related mechanisms such as recrystallization and grain growth, 
the prediction of deformation texture is also sensitive to the used 
homogenization scheme \cite{Castelnau-et-al-2009}. 
Regarding the transition zone, texture development associated with large shear 
strain have been investigated in \cite{Tommasi-et-al-2004} for wadsleyite using 
an earlier extension of the SC scheme (the tangent approach of  
\cite{Lebensohn-and-Tome-1993}) whose inconsistencies have been described in 
\cite{Gilormini-1995b,Gilormini-1995a} and partly corrected in \cite{Masson-et-al-2000}.
In Tommasi \textit{et al.} paper \cite{Tommasi-et-al-2004}, the effective viscosity of wadsleyite and its relation with the active slip systems has not been investigated. 

The POSO version of the Self-Consistent scheme was up to recently the most 
accurate mean-field method for predicting the effective viscoplastic behavior 
of highly anisotropic materials such as mantle rocks. 
As shown in e.g. \cite{Idiart-et-al-2006,Lebensohn-et-al-2011a}, this method 
provides results in very good agreement with computational homogenization 
(providing reference solutions) e.g. based on the Fourier Transform (FFT, 
sometimes also denoted spectral method) introduced by \cite{Moulinec199869}.
Among its advantages, the POSO-SC method complies with the variational upper 
bound 
\cite{Ponte-Castaneda-1991} that has been applied to highly anisotropic 
polycrystals in \cite{Liu-et-al-2003,Liu-et-al-2003b,Nebozhyn-et-al-2000}.
However, it lacks duality (stress and strain formulations lead to different 
results) and the link between the behavior of the thermo-elastic polycrystal 
used as a reference (called Linear Comparison Composite, LCC) and the 
non-linear viscoplastic behavior of the real polycrystal of interest requires 
complex computation of some corrective terms 
\cite{Idiart-and-Ponte-Castaneda-2007a} that is rarely carried out in practice.
The above mentioned limitations of POSO-SC have been ruled out recently with 
the Fully Optimized Second-Order (FOSO) method 
\cite{PonteCastaneda2015,Song2018}, which 
has, to the best of our knowledge, only been applied yet to porous sea ice 
\cite{Das2019}.

This paper focusses on the rheological behavior of $\text{Mg}_2\text{SiO}_4$ 
wadsleyite.
In section \ref{sec:2}, we provide the strength of the various slip systems 
based on computational mineral physics and the corresponding $P-T$ dependent 
constitutive relation at the grain scale. 
We present in section \ref{sec:3} how the mean-field (FOSO-SC) and full-field 
(FFT) homogenization schemes have been applied.
Results are then presented and discussed in section \ref{sec:4} and compared to 
the available 
experimental results.


\section{Viscoplastic behavior of slip systems}
\label{sec:2}
\subsection{At the dislocation scale from generalized stacking fault energies}

The modelling of gliding properties of dislocations in wadsleyite has been 
initiated by Ritterbex et al. \cite{Ritterbex20162085} for the Mg end-member of 
$\text{(Mg,Fe)}_2\text{SiO}_4$, i.e. the pure magnesian Mg$_2$SiO$_4$ 
composition.
The first step of the model is to identify the most important slip systems and to determine the specific atomic arrangements which build the dislocation core. 
Such calculations are usually performed at the atomic scale. 
However, the long range displacement fields of dislocations impose the use of 
classical molecular dynamic simulation instead of highly accurate density 
functional theory based \textit{ab-initio} calculations. 

\begin{figure}[h!]
	\centering 
	\includegraphics[width=13cm,clip]{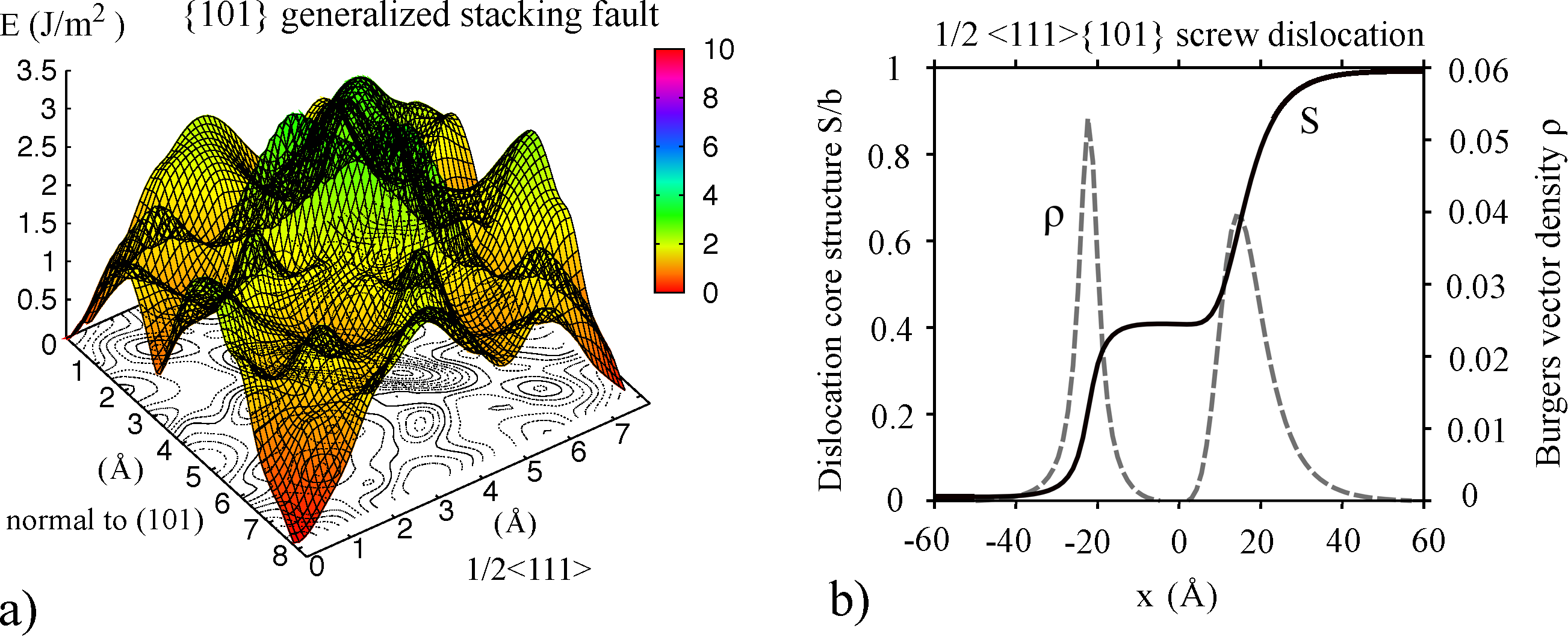}
	\caption{Dislocation modeling in $\text{Mg}_2\text{SiO}_4$ wadsleyite at 
		$15\;\giga\pascal$ (a) Generalized Stacking fault calculated by 
		\cite{Metsue20101467} for a $\{101\}$ plane. The easy shear path is 
		along 
		$<111>$. (b) Dislocation core profile for a $1/2<111>\{101\}$ 
		dislocation 
		calculated from the Peierls-Nabarro Galerkin method. $S$ is the 
		disregistry 
		profile.}
	\label{fig:gamma-surf}
\end{figure}

Another option to compute dislocation core properties is to use the  
Peierls-Nabarro model which relies on generalized stacking fault (GSF) 
energies. 
GSF's probes the ability of a perfect crystal structure to undergo a rigid body 
shear localized in a specific plane, and, combined with the Peierls-Nabarro 
approach, allows to search for potential easy slip systems of any crystalline 
materials. 
More importantly, such calculations do not require supercells containing large 
numbers of atoms so they can be performed by using \textit{ab initio} methods, 
ensuring that the effect of high pressure on bond energy or ion interactions is 
accurately taken into account. 
In orthorhombic wadsleyite (lattice parameters $a=5.70\angstrom$, 
$b=11.44\angstrom$, $c=8.26\angstrom$ at $18.5\;\giga\pascal$), the easiest 
slip systems are $[100](010)$ and $1/2<111>\{101\}$.
An example is provided in Figure \ref{fig:gamma-surf}a which shows a GSF 
calculated in a $\{101\}$ plane at $15\;\giga\pascal$. 
On the vertical axis is represented the excess energy associated with a rigid 
body shift in the considered plane. 
One can deduce that in $\{101\}$ planes, the easiest shear path is along 
$<111>$ directions. 

%

As recalled above, the GSF serves as input for the 
Peierls-Nabarro model used to compute dislocation core structures 
by minimization of the total energy of the system composed of the elastic 
strain energy and the inelastic stacking fault energy across the potential 
glide planes. 
Ritterbex et al. \cite{Ritterbex20162085} used a generalization of the 
Peierls-Nabarro model in the framework of the element-free Galerkin method 
which allows for the introduction of multiple glide planes in order to describe 
more general core structures involving spreading in several planes. 
In wadsleyite, for all slip systems considered, dislocations exhibit a planar 
core involving two well-separated partial dislocations enclosing a stacking 
fault (see an illustration in Figure \ref{fig:gamma-surf}b). 

Besides the core structure, \textit{i.e.} the atomic arrangement within the vicinity of the dislocation core, Ritterbex \textit{et al.} computed the lattice friction experienced by dislocations on each slip system. 
The amount of lattice friction is often described as the height of the Peierls 
potential or quantified through the maximum of the derivative of the potential, 
the Peierls stress. 
Nevertheless, whereas Peierls stress or Peierls potential are strictly related to the core of dislocations, the glide of dislocations at finite temperature is thermally activated. 
Thermal activation means that glide at finite temperature is controlled by the nucleation and propagation of unstable kink-pairs over the Peierls barrier, assisted by the resolved shear stress. 
Ritterbex \textit{et al.} provided a full description of these kink processes 
according to an elastic interaction model \cite{Koizumi19933483}.
For wadsleyite, since dislocations are dissociated, the kink-pair mechanism may 
involve different types of kink nucleation events depending on the stress 
regime. 
Indeed, kink nucleation must occur on both partials with sequences of events 
that can either be correlated or uncorrelated leading to distinct nucleation 
enthalpy depending on the stress regime \cite{Ritterbex2018}. 
Finally, from the kink-pair nucleation enthalpy, the dislocation velocity law 
can be formulated as a function of temperature and resolved shear stress.
The corresponding constitutive relation has to account for these two stress 
regimes, leading to two different expressions of the shear-rate on the slip 
system
\begin{align}
\label{eq:initc}
\dot\gamma_c(\tau) & = A_c\sqrt{\rho_m} \exp \left\{ -\frac{B_c}{k_b T} \left[1-\left(\frac{\tau}{\tau_p}\right)^{\alpha_c}\right]^{\beta_c} \right\} \\
\label{eq:initu}
\dot\gamma_u(\tau) & = A_u\sqrt{\rho_m} \frac{\tau-\tau_c}{\tau} \exp \left\{ -\frac{B_u}{k_b T} \left[1-\left(\frac{\tau-\tau_c}{\tau_p}\right)^{\alpha_u}\right]^{\beta_u} \right\}
\end{align}
where indexes $c$ and $u$ stand for correlated and uncorrelated kink-pair 
nucleation regimes, respectively.
Here, $\rho_m$ is the density of mobile dislocations, $T$ is the temperature, 
$\tau_p$ is the Peierls stress, $A$ and $B$ are two constants, and $k_b\simeq 
1.3806 \times 10^{-23} \meter^2\; \kilo\gram\; \second^{-2}\; \kelvin^{-1}$ is 
the Boltzmann constant.
$A$ is related to the shear-rate at $\tau=\tau_c$, marking the transition 
stress between correlated and uncorrelated regimes,  whereas $B$ is an 
activation energy.
The resolved shear stress $\tau$ acting on the slip system is given by 
\begin{equation}
\label{eq:tau}
\tau=\mb{S}:\bs{\sigma}
\end{equation}
where $\mb{S}$ is the Schmid tensor (see appendix \ref{sec:app1}), $\bs{\sigma}$ the local deviatoric stress tensor, and '$:$' the twice contracted product.
The resulting shear-rate on a given slip system is given by
\begin{equation}
\label{eq:gammac}
\dot\gamma(\tau) = \dot\gamma_c(\tau)
\end{equation}
when $\tau \leq \tau_c$ and 
\begin{equation}
\label{eq:gammacu}
\dot\gamma(\tau) = \frac{1}{2} \left[\dot\gamma_c(\tau)+\dot\gamma_u(\tau)\right] 
\end{equation}
otherwise.
For waldsleyite, the modelling that has been performed for dislocations with 
screw character which are the rate-limiting ones. 
The easiest slips include the family $1/2<111>\{101\}$ with four individual 
slip systems ($1/2[11\bar 1](101)$, $1/2[1\bar 1\bar1](101)$, $1/2[111](10\bar 
1)$, $1/2[1\bar 11](10\bar 1)$), and the additional system $[100](010)$.
At stresses higher than $\tau_c=455\;\mega\pascal$ as considered here, the rheology of $1/2<111>\{101\}$ screw dislocations at $15\;\giga\pascal$ is governed by equation (\ref{eq:gammacu}) whereas that for $[100](010)$ is given by (\ref{eq:gammac}).
The corresponding rheological coefficients are indicated in Table 
\ref{tab:rheol_coef}.
Finally, the local strain-rate $\dot{\varepsilon}$ at position $\mb{x}$ inside 
a grain, resulting from the glide of dislocations on all slip systems, is 
given by
\begin{equation}
\dot{\varepsilon}(\mb{x}) = \sum_{s=1}^S \mb{S}_{(s)} \dot\gamma_{(s)}(\mb{x})
\label{eq:loc_constit_rel}
\end{equation}
with $\mb{S}_{(s)}$ the Schmid tensor for system $(s)$ and $S$ ($=5$) the 
number of slip systems.
Note that the four systems $1/2<111>\{101\}$ are independent and do not allow axial strain of the crystal lattice along the lattice direction $\mb{b}$ (see appendix \ref{sec:app1}).
The fifth system $[100](010)$ does not add any additional degree of freedom, and therefore the wadsleyite crystal is left with only four independent slip systems. 
As discussed in \cite{Hutchinson1977,Detrez-et-al-JMPS-2015}, four systems are sufficient to accommodate locally any viscoplastic deformation of the polycrystalline aggregate.

\begin{table}[h!]
	\begin{center}	
		\begin{tabular}{|c|c|c|c|c|c|c|c|c|c|c|}
			\hline
			& $\tau_c$ & $\tau_p$ & $A_c$ & $B_c$ & $\alpha_c$ & $\beta_c$ & $A_u$ & $B_u$ & $\alpha_u$ & $\beta_u$ \\
			\hline
			$1/2<111>\{101\}$ & 455 & 3500 & 2190 & $1.97\times 10^{-18}$ & 0.5 & 1.6 & 4380 & $8.49\times 10^{-19}$ & 1.0 & 5.0 \\
			$[100](010)$     &      & 4800 & 1208 & $2.0\times 10^{-18}$  & 0.5 & 1.03 &  &  &  &  \\
			\hline
		\end{tabular}
		\caption{Rheological parameters of $1/2<110>\{110\}$ and $[100](010)$ 
			screw dislocations in wadsleyite at $15\;\giga\pascal$ for 
			$\tau\geq\tau_c$. Units are $\meter.\second^{-1}$ for $A$, $\joule$ for 
			$B$, and $\mega\pascal$ for $\tau$.}
		\label{tab:rheol_coef}
	\end{center}
\end{table}

\subsection{Power law approximation of slip system behavior}

Although homogenization models could in principle be applied with constitutive relations having an exponential form as (\ref{eq:gammac}) and (\ref{eq:gammacu}), it is more common to use instead a power-law behavior
\begin{equation}
\label{eq:powerlaw}
\dot\gamma_{pl} = \dot\gamma_0 \left| \frac{\tau}{\tau_0}\right|^{n-1} \frac{\tau}{\tau_0} 
\end{equation}
where the index $pl$ stands for power-law.
This is the form commonly implemented in homogenization codes.
Here, $\dot\gamma_0$ and $\tau_0$ are two constants to be determined, a reference shear-rate and a reference shear stress, respectively.
We consider here the value $\dot\gamma_0 = 10^{-5}\second^{-1}$ as a typical 
strain-rate achieved during laboratory mechanical tests.
Form (\ref{eq:powerlaw}) can be seen as a power-law approximation of (\ref{eq:gammac}) or (\ref{eq:gammacu}) valid in a range of $\tau$ values close to a given reference value denoted $\tau_r$.
To find $\tau_0$, one expresses that the power-law approximation is tangent to the exponential form at $\tau=\tau_r$: 
\begin{equation}
\dot\gamma_{pl}(\tau_r) = \dot\gamma(\tau_r) \; , \qquad
\left.\frac{\partial\dot\gamma_{pl}}{\partial\tau}\right|_{\tau=\tau_r}  = \left.\frac{\partial\dot\gamma}{\partial\tau}\right|_{\tau=\tau_r} \;.
\end{equation}
This leads to the following expression for the stress exponent $n$ 
\begin{equation}
\label{eq:expr_n}
n= \left. \frac{\partial \log \dot\gamma}{\partial \log\tau} \right|_{\tau=\tau_r}
=\frac{\tau_r}{\dot\gamma(\tau_r)} \left. \frac{\partial \dot\gamma}{\partial\tau} \right|_{\tau=\tau_r}
\end{equation}
while the reference stress $\tau_0$ reads
\begin{equation}
\tau_0 = \tau_r\left(\frac{\dot\gamma(\tau_r)}{\dot\gamma_0}\right)^{-1/n} \;.
\end{equation}
These can be calculated using the following expressions for the derivatives in (\ref{eq:expr_n})
\begin{align}
\frac{\partial\dot\gamma_c}{\partial\tau} = &
- a \exp \left\{ b \left[1-\left(\frac{\tau}{\tau_p}\right)^\alpha\right]^\beta \right\}
\frac{\alpha\beta b}{\tau}
\left(\frac{\tau}{\tau_p}\right)^{\alpha}
\left( 1-\left( \frac{\tau}{\tau_p}\right)^\alpha\right)^{\beta-1} 
\\
\frac{\partial\dot\gamma_u}{\partial\tau} = &
a \exp \left\{ b \left[1-\left(\frac{\tau-\tau_c}{\tau_p}\right)^\alpha\right]^\beta \right\}
\left\{
\frac{\tau_c}{\tau^2} -  \frac{\alpha\beta b}{\tau}
\left(\frac{\tau-\tau_c}{\tau_p}\right)^{\alpha}
\left( 1-\left( \frac{\tau-\tau_c}{\tau_p}\right)^\alpha\right)^{\beta-1}
\right\} 
\end{align}
with $a=A\sqrt{\rho_m}$, $b=-{B}/(k_b T)$, and with indexes $c$ and $u$ left for the sake of clarity.

\begin{figure}[h!]
	\centering \includegraphics[width=6cm]{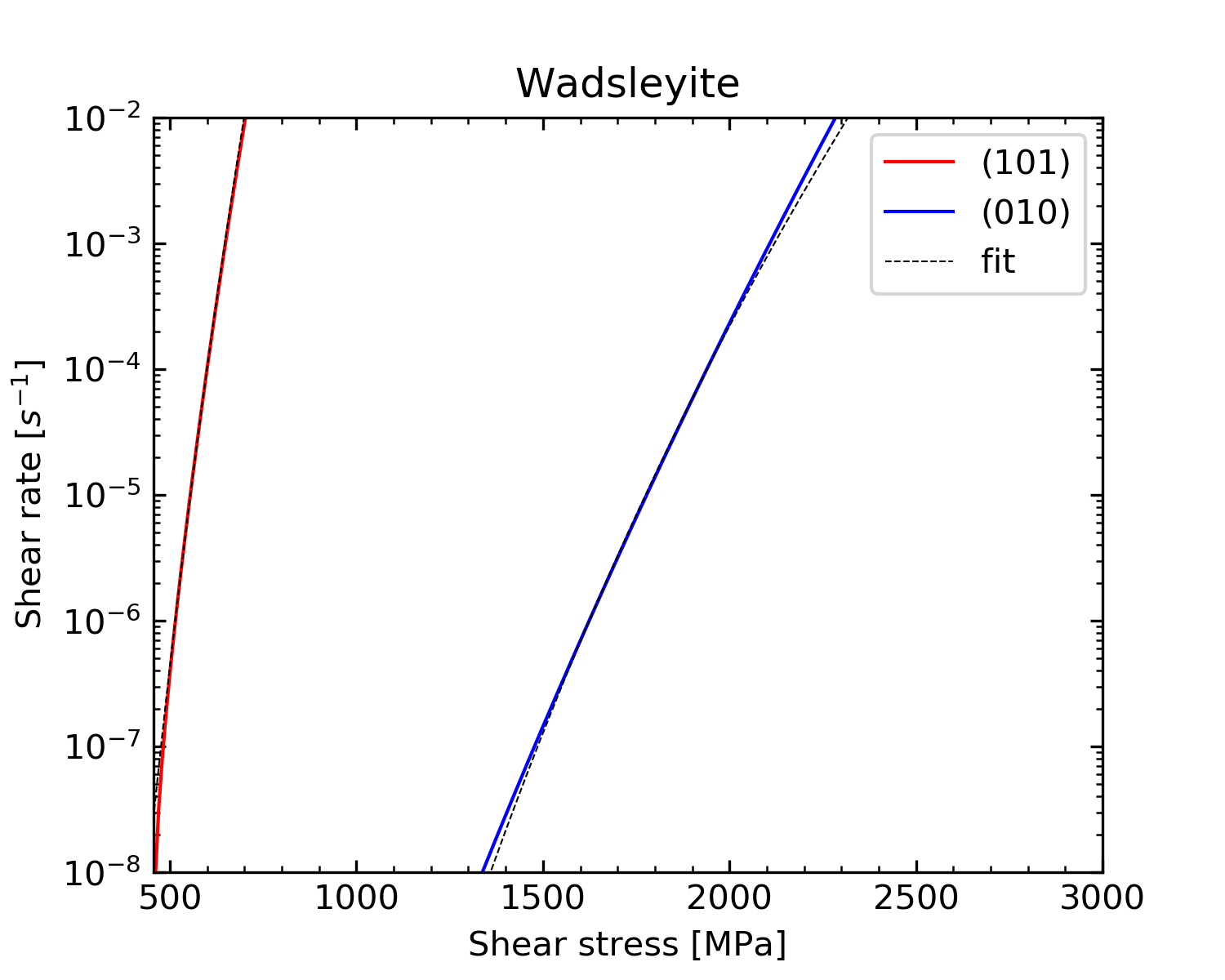}
	\centering \includegraphics[width=6cm]{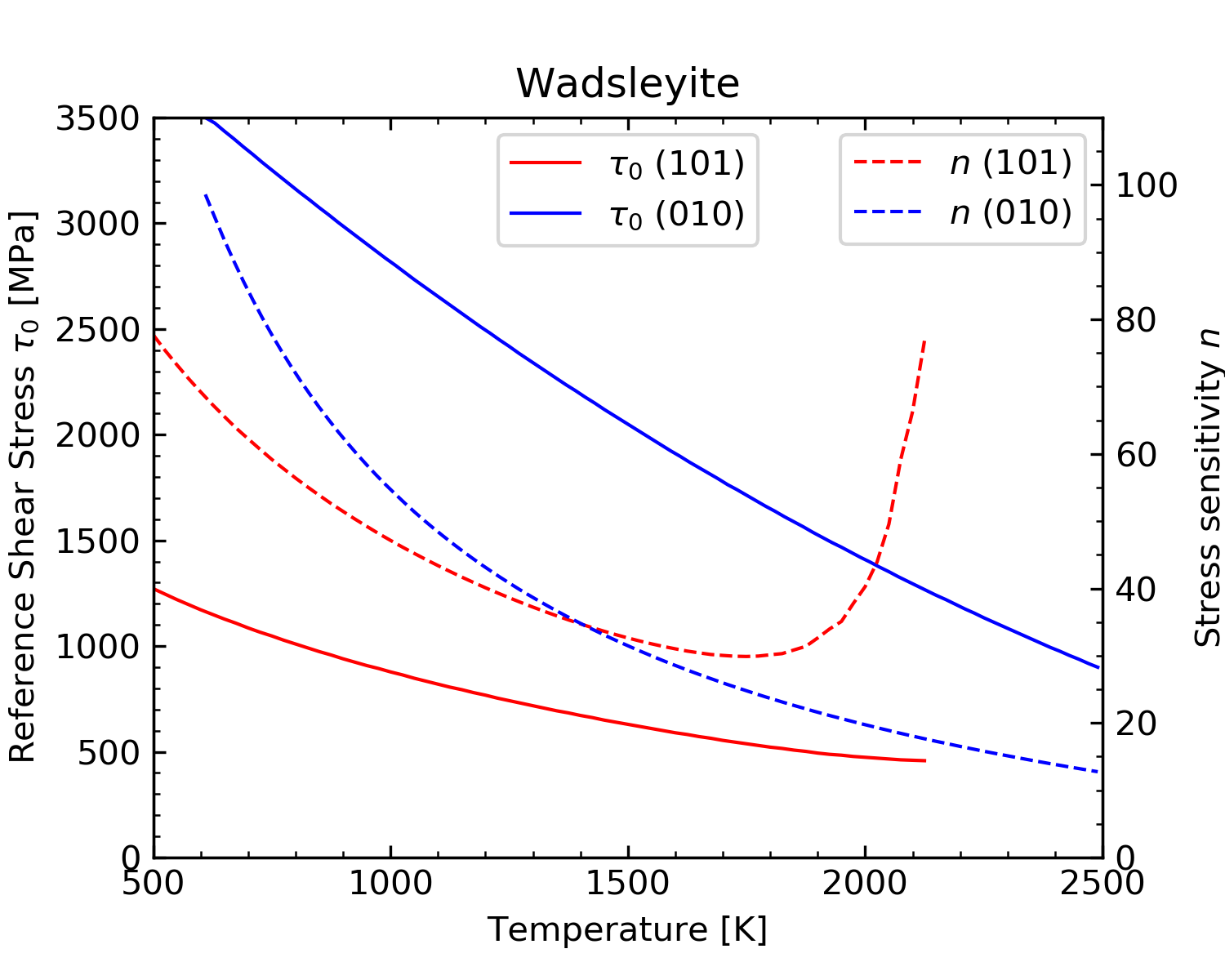}
	\caption{(left) Power-law approximations of the slip system behaviors for 
		wadsleyite at $\dot\gamma=10^{-5}\second^{-1}$: comparison of the original 
		exponential constitutive relation with its power-law approximation (fit) at 
		$1700\kelvin$. (right) Evolution of $\tau_0$ and $n$ of the power-law 
		approximations of the slip system behavior at 
		$\dot\gamma=10^{-5}\second^{-1}$, for various temperatures.} 
	\label{fig:powerlawFit-Wads}
\end{figure}

Power-law approximations of the exponential constitutive relations have been 
calculated under typical conditions corresponding to high $P$, $T$ laboratory 
mechanical tests, \textit{i.e.} considering a density of mobile dislocations 
$\rho_m=10^{12}\;\meter^{-2}$ and a strain-rate $\dot\gamma=\dot\gamma_0$ 
(leading to $\tau_0=\tau_r$).
It is observed that the power-law provides a very good approximation of the 
original exponential behavior, as illustrated in Figure 
\ref{fig:powerlawFit-Wads}a at $1700\kelvin$. 
The resulting power-law parameters are provided in Figure \ref{fig:powerlawFit-Wads}b for a large temperature range.
It turns out that the rheology of the various slip systems is strongly 
non-linear.
Values of $n$ are between $\sim17$ and $100$ for the considered 
temperature range, with a different behavior for both slip systems.
Moreover, the reference shear stress for the system $[100](010)$ is 
significantly stiffer 
($\sim 3$ times) than for $1/2<111>\{101\}$.

\section{Polycrystal Modeling}
\label{sec:3}

We now proceed to the next scale transition, \textit{i.e}. estimation of the 
effective (average) viscoplastic behavior of a representative polycrystalline 
aggregate composed of a large number of grains.
We consider the case for which grains are randomly oriented (random 
crystallographic texture) and exhibit equiaxed shapes so that the effective 
behavior can be considered as isotropic.
In the sequel, effective (or homogenized) quantities are denoted with a tilde ( 
$\tilde.$ ) and volume average ones with a bar ( $\bar.$ ).
The effective viscoplastic behavior thus reads
\begin{equation}
\label{eq:effective-behavior}
\dot{\bar\varepsilon}_{eq} = \dot\gamma_0 \left( \frac{\bar\sigma_{eq}}{\tilde\sigma_0} \right)^{\tilde n}
\end{equation}
where $\bar\sigma_{eq}$ is the effective von Mises equivalent stress 
($\bar\sigma_{eq}=\sqrt{\frac{3}{2} \bar\sigma_{ij}\bar\sigma_{ij}}$), 
$\dot{\bar\varepsilon}_{eq}$ is the effective von Mises equivalent 
strain-rate ($\dot{\bar\varepsilon}_{eq}=\sqrt{\frac{2}{3} 
	\dot{\bar\varepsilon}_{ij}\dot{\bar\varepsilon}_{ij}}$), where 
$\bar{\bs{\sigma}}=<\bs{\sigma}(\mb{x})>$ and 
$\dot{\bar{\bs{\varepsilon}}}=<\dot{\bs{\varepsilon}}(\mb{x})>$ are the mean 
deviatoric stress and strain-rate tensors respectively.

To estimate the effective reference stress $\tilde\sigma_0$ and effective 
stress sensitivity $\tilde n$, and to investigate how the stress/strain-rate  
are distributed within grains, we make use of two scale transition methods, the 
mean-field FOSO-SC scheme recently proposed by \cite{Song2018} and the 
full-field computational homogenization based on the FFT method 
\cite{Moulinec199869}.
All results below have been obtained for uniaxial deformation under a 
prescribed macroscopic strain-rate $\dot{\bar\varepsilon}_{eq}=\dot\gamma_0$, 
so that $\bar\sigma_{eq} = \tilde\sigma_0$.

\subsection{Mean-field approach - FOSO-SC scheme}
\label{sec:mean-field}

Mean-Field homogenization of heterogeneous materials is relatively 
straightforward when the mechanical behavior of each mechanical phase 
(\textit{i.e.} 
grains with given orientation) is linear and homogeneous.
This is the case for linear thermo-elastic polycrystalline aggregates where the 
local 
constitutive relation for a phase $p$ reads
\begin{equation}
\label{eq:local}
\bs{\varepsilon}(\mb{x}) = \mb{M}^{(p)}:\bs{\sigma}(\mb{x}) + 
\bs{\varepsilon}_0^{(p)} 
\end{equation}
where $\mb{M}^{(p)}$ and $\bs{\varepsilon}_0^{(p)} $ only depend on the crystal 
orientation of the considered grain.
This leads to an effective behavior of the same form
\begin{equation}
\label{eq:effective}
\bar{\bs{\varepsilon}} = \tilde{\mb{M}}:\bar{\bs{\sigma}} + 
\tilde{\bs{\varepsilon}}_0 \;.
\end{equation}
In that case, to estimate the effective compliance given by
\begin{equation}
\tilde{\mb{M}}=<\mb{M}(\mb{x}):\mb{B}(\mb{x})> \;,
\label{eq:Mtilde}
\end{equation}
with $\mb{B}$ the stress concentration tensor of the purely elastic problem 
(i.e. when $\bs{\varepsilon}_0^{(p)}=\mb{0}$) defined as 
$\bs{\sigma}(\mb{x})=\mb{B}(\mb{x}):\bar{\bs{\sigma}}$, 
it is sufficient to estimate the phase-average of $\mb{B}$, denoted 
$<\mb{B}>^{(p)}$.
Indeed, the volume integral in (\ref{eq:Mtilde}) can be transformed into a 
discrete sum over all mechanical phases 
\begin{equation}
\tilde{\mb{M}} = \sum_p f_p \mb{M}^{(p)}:<\mb{B}(\mb{x})>^{(p)} \;,
\label{eq:Mtilde_TE}
\end{equation}
with $f_p$ indicating the volume fraction.
This is why mean-field homogenization is very efficient numerically.

The situation for non-linear polycrystals as considered here is more complex as 
the local compliance $\mb{M}(\mb{x})$ defined after (\ref{eq:loc_constit_rel}) 
depends on the local stress state which is itself heterogeneously distributed 
within deforming grains \cite{Ponte-Castaneda-and-Suquet-1998}.
The solution (\ref{eq:Mtilde_TE}) therefore cannot be used directly.
The standard approach to address this issue relies on a linearization of the 
non-linear polycrystal of interest in order to end up with a 
thermo-elastic-like 
material (LCC), exhibiting the same microstructure as the original non-linear 
one, 
but for which the standard thermo-elastic homogenized solution applies.
Finding the proper linearization procedure is a difficult task, this is why 
several propositions can be found in the literature (some of them are listed 
below).
The exact link between the effective behavior of the non-linear polycrystal and 
of the linearized thermo-elastic one is another difficulty that has often been 
left aside \cite{Idiart-and-Ponte-Castaneda-2007a}.
The impact of the used linearization procedure on the effective behavior 
becomes critical for highly non-linear materials or, equivalently, highly 
anisotropic local behavior or high mechanical contrast between the phases.
As seen in the previous section, waldseyite exhibits both very high 
non-linearity and anisotropy, which constitutes a challenge for mean-field 
homogenization.

The most advanced linearization procedure nowadays, that we will use here, is 
the Fully Optimized Second Order (FOSO) scheme proposed recently by 
\cite{Song2018}.
In short, starting from a variational formulation of the problem, the optimal 
linearization, leading to the definition of the linear thermo-elastic 
comparison material, is defined as an optimization problem.
Compared to the Partially Optimized (POSO) formulation \cite{Ponte2002a737}, 
full optimization could be carried out in FOSO, hence the name.
This formulation has several significant advantages compared to the previous 
POSO one: 
(i) there is no duality gap, \textit{i.e.} stress and strain-rate formulations 
yield similar results,
(ii) stress and strain-rate field statistics in the linearized and non-linear 
polycrystals are identical, there is no need to compute the corrective terms 
as in \cite{Idiart-and-Ponte-Castaneda-2007a}.
As POSO, the FOSO approach complies by construction with the known upper bounds 
for the effective behavior.
First application of FOSO-SC to viscoplastic porous hexagonal polycrystals 
(similar to sea ice) yields a rheology in very good agreement with the 
reference results obtained by the FFT full-field homogenization \cite{Das2019}.
In the FOSO-SC approach, the local behavior of the linear thermo-elastic 
comparison polycrystal reads, at the slip system level,  
\begin{equation}
\label{eq:lin1}
\dot\gamma_{(s)}(\mb{x}) = m_{(s)}^{(p)} \tau_{(s)}(\mb{x})  + 
\dot e_{(s)}^{(p)} 
\end{equation}
where the compliance $m_{(s)}^{(p)}$ and stress-free 
strain-rate $\dot e_{(s)}^{(p)}$ depend on both the first moment 
$\bar\tau_{(s)}^{(p)} = <\tau_{(s)}(\mb{x})>^{(p)}$ and second moment 
$\bar{\bar\tau}_{(s)}^{(p)} = <\tau_{(s)}^2(\mb{x})>^{(p)}$ 
of 
resolved shear stress in the mechanical phase $p$:
\begin{equation}
\label{eq:lin2}
m_{(s)}^{(p)} = \frac{\dot\gamma(\hat\tau) - 
	\dot\gamma(\check\tau)}{\hat\tau - \check\tau} \;, \qquad 
\dot e_{(s)}^{(p)} = \dot\gamma(\check\tau) - m_{(s)}^{(p)} \check\tau
\end{equation}
with $\check\tau$ and $\hat\tau$ defined as
\begin{equation}
\check\tau = 
\bar\tau-\sqrt{\frac{1-\alpha}{\alpha}}\sqrt{\bar{\bar\tau}-\bar\tau^2} 
\;\text{sign}(\bar\tau) \;,
\qquad
\hat\tau = 
\bar\tau+\sqrt{\frac{1-\alpha}{\alpha}}\sqrt{\bar{\bar\tau}-\bar\tau^2} 
\;\text{sign}(\bar\tau) \;.
\label{eq:tau_hatcheck}
\end{equation}
In (\ref{eq:tau_hatcheck}), $\sqrt{\bar{\bar\tau}-\bar\tau^2}$ is the standard 
deviation of the shear stress distribution acting on the slip system. 
Using the value $\alpha=0.5$ recommended in \cite{Song2018}, $\check\tau$ and 
$\hat\tau$ are therefore one standard deviation below and above the mean 
value $\bar\tau$.

In comparison, the POSO formulation can be obtained using $\check\tau = 
\bar\tau$ and the same definition for $\check\tau$ as in 
(\ref{eq:tau_hatcheck}) while the variational upper bound of 
\cite{Ponte-Castaneda-1991} corresponds to $\check\tau=0$ and 
$\hat\tau=(\bar{\bar\tau})^{1/2}$.
On the other hand, the earlier affine scheme 
\cite{Masson-et-al-2000}, which is not based on a variational formulation, 
would be obtained by taking the limit of (\ref{eq:lin2}) for 
$\alpha \rightarrow 1$, therefore not making use of the shear stress 
fluctuation 
at 
the slip system level.
The even earlier tangent formulation of \cite{Lebensohn-and-Tome-1993} applied 
to wadsleyite in \cite{Tommasi-et-al-2004} is based on a similar formulation as 
the affine one but with a compliance divided by $n$ and $\dot e_{(s)}^{(p)}=0$.

Application to wadsleyite has been performed considering equiaxe grain shapes 
and using a 
set of 2000 crystal orientations generated by a Sobol quasi-random sequence, 
that provides better overall isotropy than the random orientation usually 
chosen.
As already mentioned, wadsleyite exhibits highly anisotropic and non-linear 
behavior at the grain scale, and this makes the numerical convergence of the 
FOSO-SC model rather delicate. 
The used numerical procedure is detailed in appendix \ref{sec:app2}.
It is worth noting that the model converges without having to introduce any 
additional unphysical slip systems often used in the literature to reach a 
total of 5 independent slip systems.

\subsection{Full-Field approach - FFT numerical homogenization}

The FFT method \cite{Moulinec199869} relies on the 3D microstructure of the 
considered polycrystal, 
which constitutes the unit cell, submitted to periodic boundary conditions. 
This unit cell is discretized into $N_{1}\times N_{2}\times N_{3}$ voxels. 
This discretization determines a regular grid in the cartesian space ${x_{d}}$ 
and a corresponding grid in the Fourier space ${\xi_{d}}$. 
The heterogeneous problem of a polycrystal exhibiting a different compliance 
$\mb{M}(\mb{x})$ at each position $(\mb{x})$ is rewritten equivalently as a 
homogeneous problem with an arbitrary homogeneous compliance  $\mb{M}^0$ and an 
additional unknown stress-free strain-rate (or polarization) field.
The solution is given by a convolution of the Green tensor associated to 
$\mb{M}^0$ with the polarization field of interest.
In the Fourier space, this convolution turns into a direct product, hence the 
very high numerical efficiency of the method.
An iterative scheme must be implemented to obtain, upon convergence, the 
compatible strain-rate field associated to the balanced stress field for nonlinear rheology, as detailed in \cite{Suquet201264}.
This FFT method provides the 'exact' solution (apart purely numerical errors) 
for the considered microstructure, but requires significantly more computing 
ressources than mean-field estimations.
Another advantage of such full-field homogenization is that details of stress 
and strain-rate distributions within the microstructure can be obtained.

\begin{figure}[h!]
	\centering \includegraphics[width=6cm,clip]{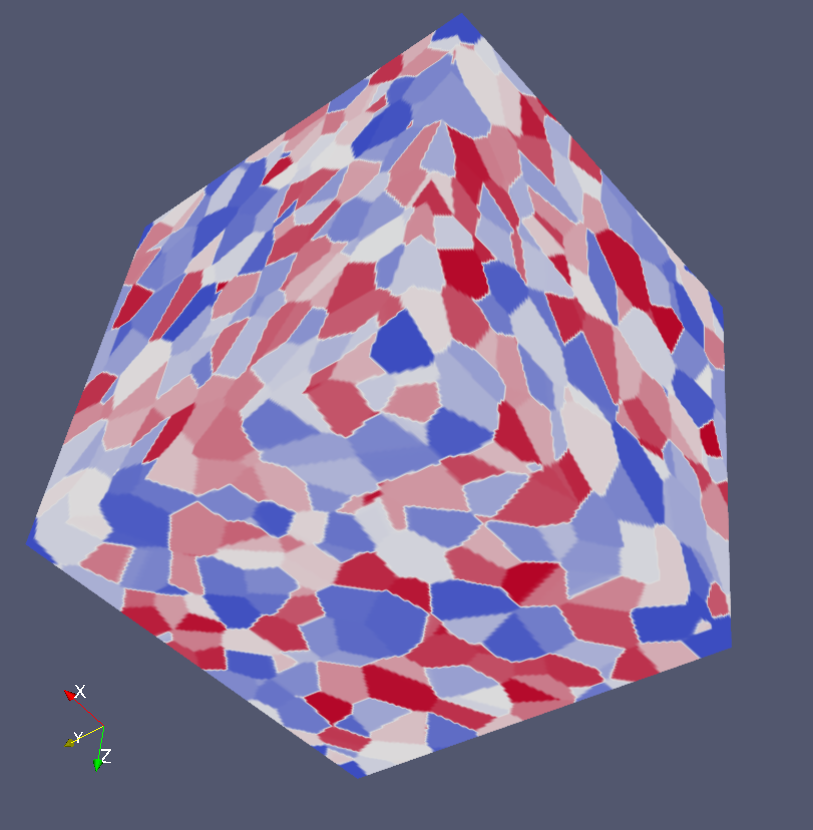}
	\caption{Typical periodic microstructure considered for FFT computations.}
	\label{fig:micro-256}
\end{figure}

For the application to wadsleyite, the considered microstructure is a 
periodical three-dimensional unit cells randomly generated by Voronoi 
tesselation, and containing 1000 grains (figure \ref{fig:micro-256}). 
Crystal orientations are chosen according to the Sobol sequence introduced in 
\ref{sec:mean-field}, leading to an effective behavior close to isotropic.
The unit cell was discretized into $256\times256\times256$ voxels.
The effective rheology has been obtained by averaging the model output for 10 
random realizations of such synthetic polycrystalline agregates.
The relative statistical uncertainty of results given below has been estimated 
according to the method proposed in \cite{Kanit-et-al-2003}. 
For example, we found that using 10 random realizations leads to an error of 
only $0.1\%$ on the effective stress $\tilde\sigma_0$.

\section{Results and discussion}
\label{sec:4}

Results provided by the 3-scales homogenization approach described above, from 
the nanometer $P-T$ dependent dislocation core structure up to the sub-meter 
polycrystal scale, is now given for a typical strain-rate corresponding to 
laboratory mechanical tests.
The reference FFT solution and the FOSO-SC estimate integrate the rheology of 
individual slip systems obtained by the Peierls-Nabarro model described in 
section \ref{sec:2}.

\begin{figure}[h!]
	\centering \includegraphics[width=10cm,clip]{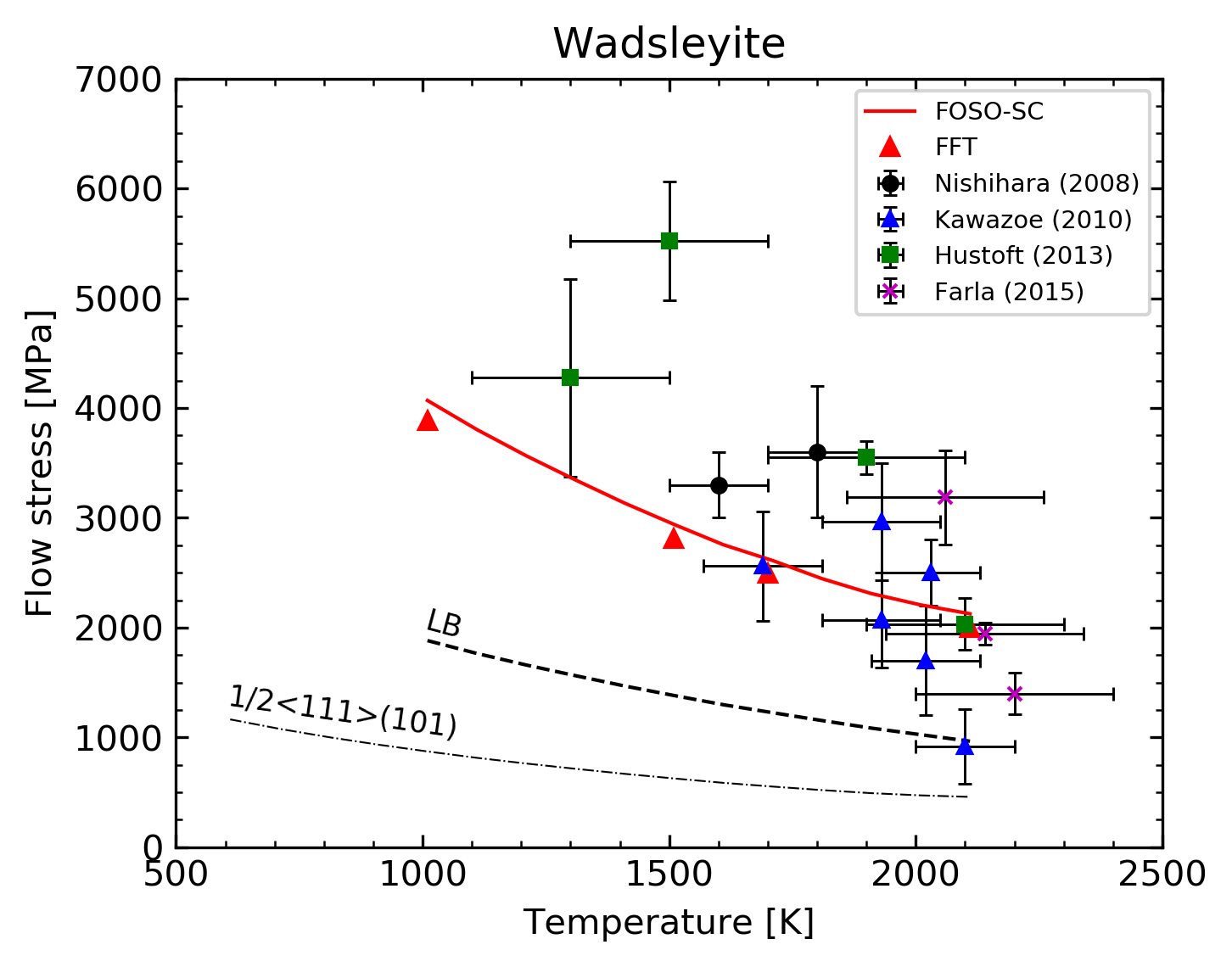}
	\caption{Macroscopic response $\tilde\sigma_0$ of wadsleyite for various 
		temperatures. Predictions of FOSO-SC and FFT polycrystal models 
		including dislocation mobility based on the GSF model are compared to 
		experimental data. The reponse of individual slip on $1/2<111>(101)$   
		($\tau$ value for $\dot\gamma=10^{-5}\second^{-1}$ as in figure 
		\ref{fig:powerlawFit-Wads}) as well as the 
		static Lower Bound (LB) are shown for comparison.}
	\label{fig:resultats-macro}
\end{figure}

Figure \ref{fig:resultats-macro} shows the flow stress $\tilde\sigma_0$ for 
temperatures ranging between $1100\kelvin$ and $2100\kelvin$.
First of all, it can be observed that the mean-field homogeization scheme 
provides an estimation of the effective behavior that lies very close to the 
reference FFT numerical solution. 
Both estimations differ by less than $5\%$ for the four temperatures computed 
by FFT (see numerical values in Table \ref{tab:various_hnl} for $1700\kelvin$).
The FOSO-SC model therefore does an excellent job considering the very strong 
viscoplastic anisotropy of wadsleyite crystals (recall that axial strain along 
lattice 
direction $\mb{b}$ is impossible) and non-linearity ($n\simeq 30$ at 
$1700\kelvin$).
The effective stress $\tilde\sigma_0$ is found to be $\sim 4.7$ times larger than the flow stress of $1/2<111>(101)$ for the whole temperature range considered.

Concerning the effective stress sensitivity $\tilde n$, it is normally related 
with those of individual slip systems in a complex way due to the mechanical 
interaction between the grains during deformation.
The situation for wadsleyite is somehow simplified as both FFT and FOSO-SC 
approaches predict no activation at all of system $[100](010)$.
In \cite{Detrez-et-al-JMPS-2015}, it has however been shown for olivine that 
even when 
the activity of accommodation mechanisms could be considered as negligeable, 
their influence on the effective stress sensitivity can be significant. 
For example, $\tilde n$ during the creep of olivine can hardly exceed a value 
of 2 when dislocation glide 
with $n=3.5$ is accommodated by a linear mechanisms such as Nabarro-Herring or 
Coble diffusion.
Olivine however exhibits only 3 independent slip systems, compared to 4 systems
for wadsleyite.
Here, slip on $[100](010)$ does not relax any kinematic constraint for grains 
deforming with the four systems of the $1/2<111>\{101\}$ family.
Consequently, it is observed that system $[100](010)$ has no effect on  
the effective stress sensitivity, and we find numerically that $\tilde n$ for 
wadsleyite is exactly equal to the $n$ value of $1/2<111>\{101\}$ slip, already 
indicated in figure \ref{fig:powerlawFit-Wads}: it lies between $18$ and $50$ 
in the considered temperature range.

\begin{figure}[h!]
	\centering 
	\includegraphics[width=14cm,clip]{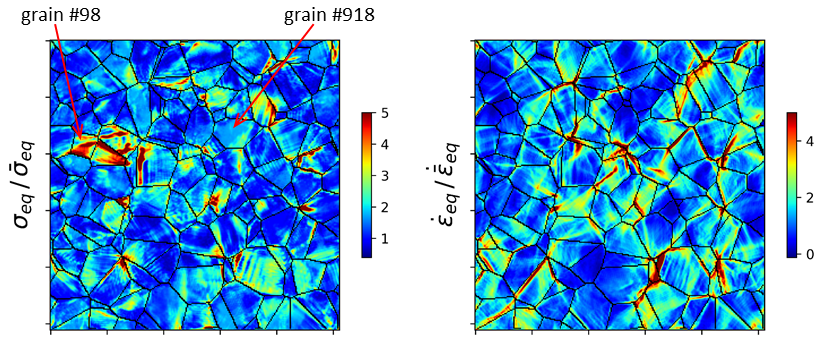}
	\caption{Distribution of (left) normalized equivalent stress 
		$\sigma_{eq}/{\bar\sigma}_{eq}$ and (right) normalized equivalent 
		strain-rate $\dot\varepsilon_{eq}/\dot{\bar\varepsilon}_{eq}$ in a 2D 
		section of a 3D FFT microstructure for wadsleyite at $1700\kelvin$.}
	\label{fig:wad-fields}
\end{figure}

\begin{figure}[h!]
	\centering 
	\includegraphics[width=6cm,clip]{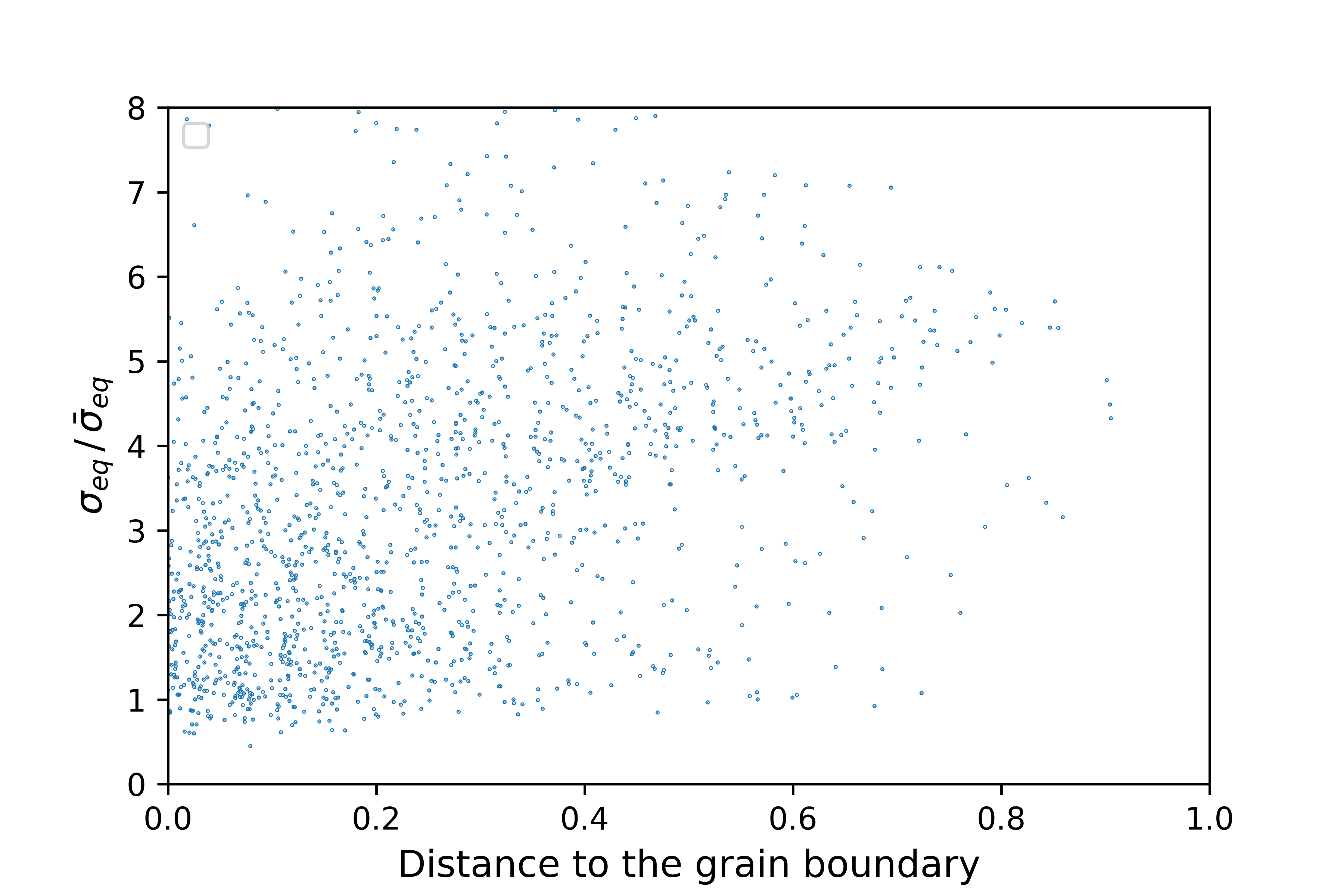}
	\includegraphics[width=6cm,clip]{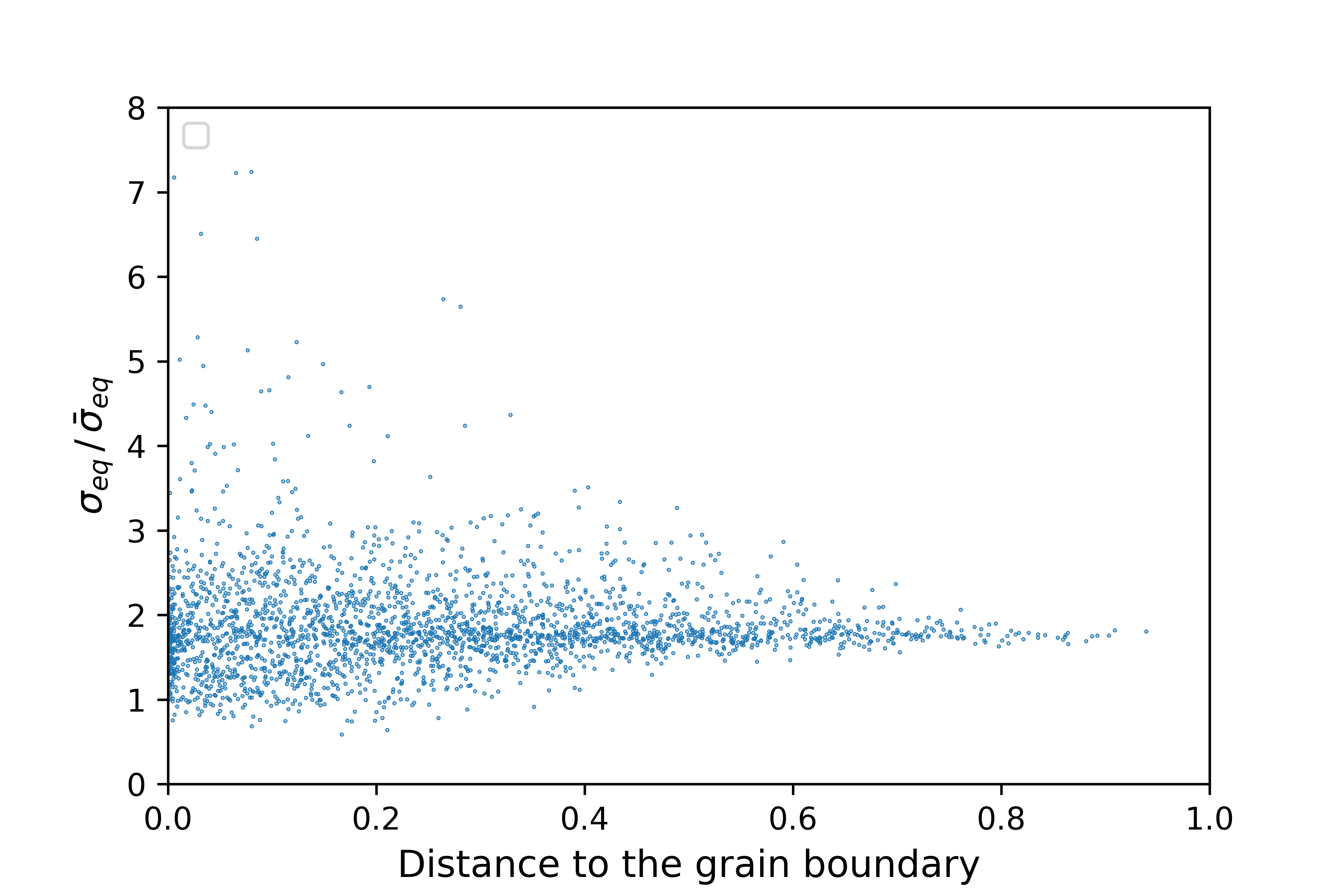}
	\caption{Effect of the distance (normalized by the grain size) to the closest grain boundary on 
		the normalized equivalent stress for (left) grain $\#98$ and (right) grain $\#918$ indicated in 
		figure \ref{fig:wad-fields}. Each dot correspond to a voxel of the grain.}
	\label{fig:wad-distance}
\end{figure}

The fields of normalized equivalent stress 
$\sigma_{eq}/{\bar\sigma}_{eq}$ and strain-rate 
$\dot\varepsilon_{eq}/\dot{\bar\varepsilon}_{eq}$ within a section of a 3D 
microstructure computed by FFT at $1700\kelvin$ is shown is figure 
\ref{fig:wad-fields}.
Large fluctuations are observed between grains (intergranular heterogeneity) 
but also inside individual grains (intragranular heterogeneity).
Hot spots corresponding to high values are clearly visible, they seem to be 
essentially located at triple junctions and grain boundaries, but note that 
locations corresponding to large equivalent stresses $\sigma_{eq}$ do often not 
exhibit at the same time a large equivalent strain-rate $\dot\varepsilon_{eq}$.
In an attempt to quantify the concentration of stress at grain boundaries, we 
have computed for each voxel of the FFT microstructures the distance to the 
nearest grain boundary.
We have carefully checked the behavior of many grains, and we show in figure 
\ref{fig:wad-distance} representative results, for the two grains ($\#98$ and 
$\#918$) indicated in figure \ref{fig:wad-fields}.
One can observe that there is a slightly smaller heterogeneity of the 
equivalent stress at grain interior for grain $\#918$ compared to voxels close 
to grain boundaries, but such trend is not observed in grain $\#98$.
The same analysis was performed at the polycrystal level, accounting for all 
grains and voxels.
The results are not shown here for the sake of conciseness, but a similar trend 
is observed. 
What can be said is that the largest values of equivalent stresses are observed 
only within voxels close to grain boundaries, but this concerns only a very 
small volume fraction of the material.
The global picture is similar to that shown in figure \ref{fig:wad-distance} 
with no significantly larger stress heterogeneity close to grain boundary.
A similar analysis was carried out in \cite{Cailletaud2003} in the case of 
cubic polycrystals, and a larger effect of the distance was found, probably due 
to the fact that the plastic anisotropy at the grain scale is much smaller in 
cubic materials than for wadsleyite.
The authors also mentioned a different behavior for hexagonal polycrystals, but 
without providing further details.
A more detailed analysis of grain boundary 
effects and  their dependence with the anisotropy and non-linearity of the 
local constitutive relation would be necessary and is left for future work.

\begin{figure}[h!]
	\centering 
	\includegraphics[width=6cm,clip]{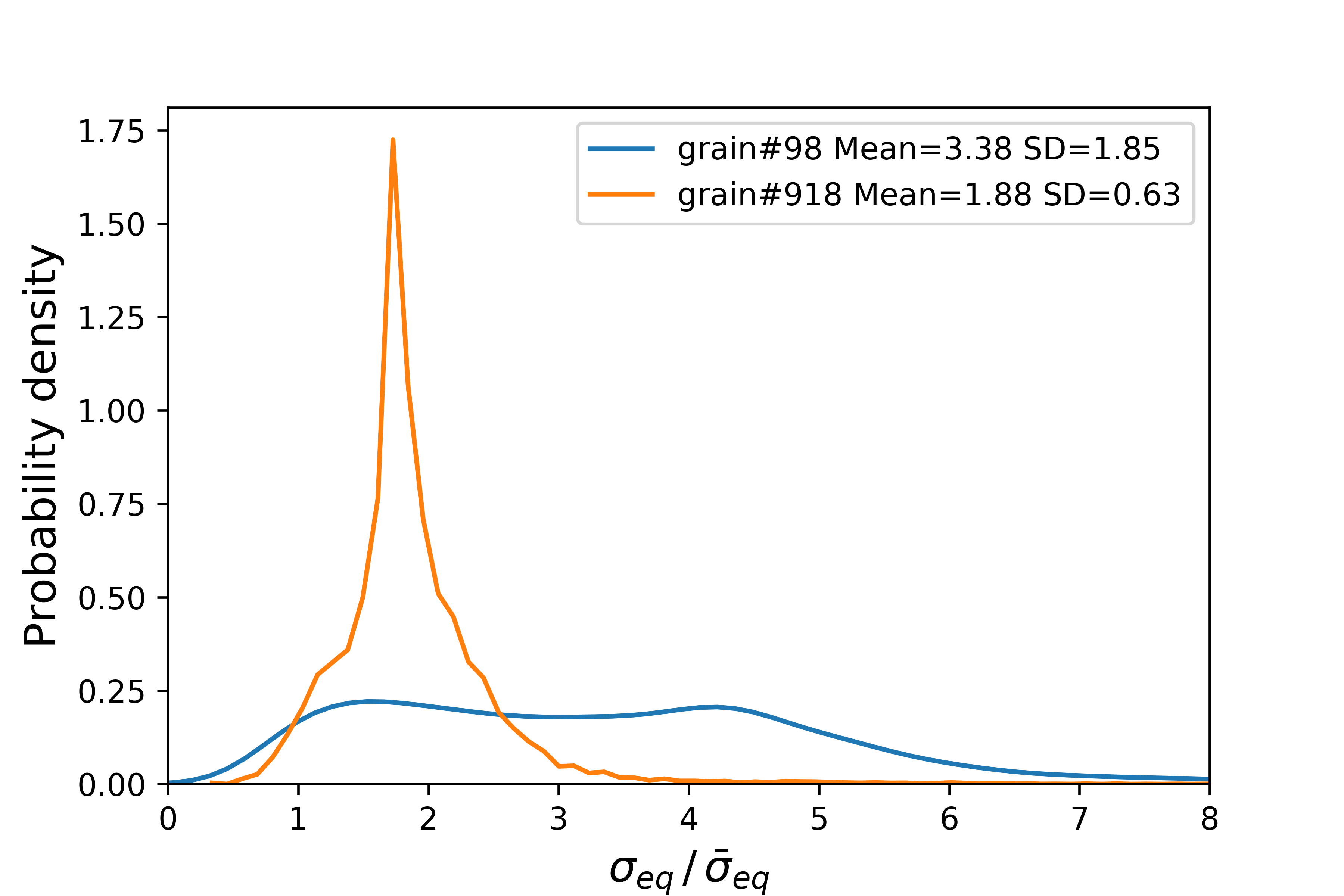}
	\includegraphics[width=6cm,clip]{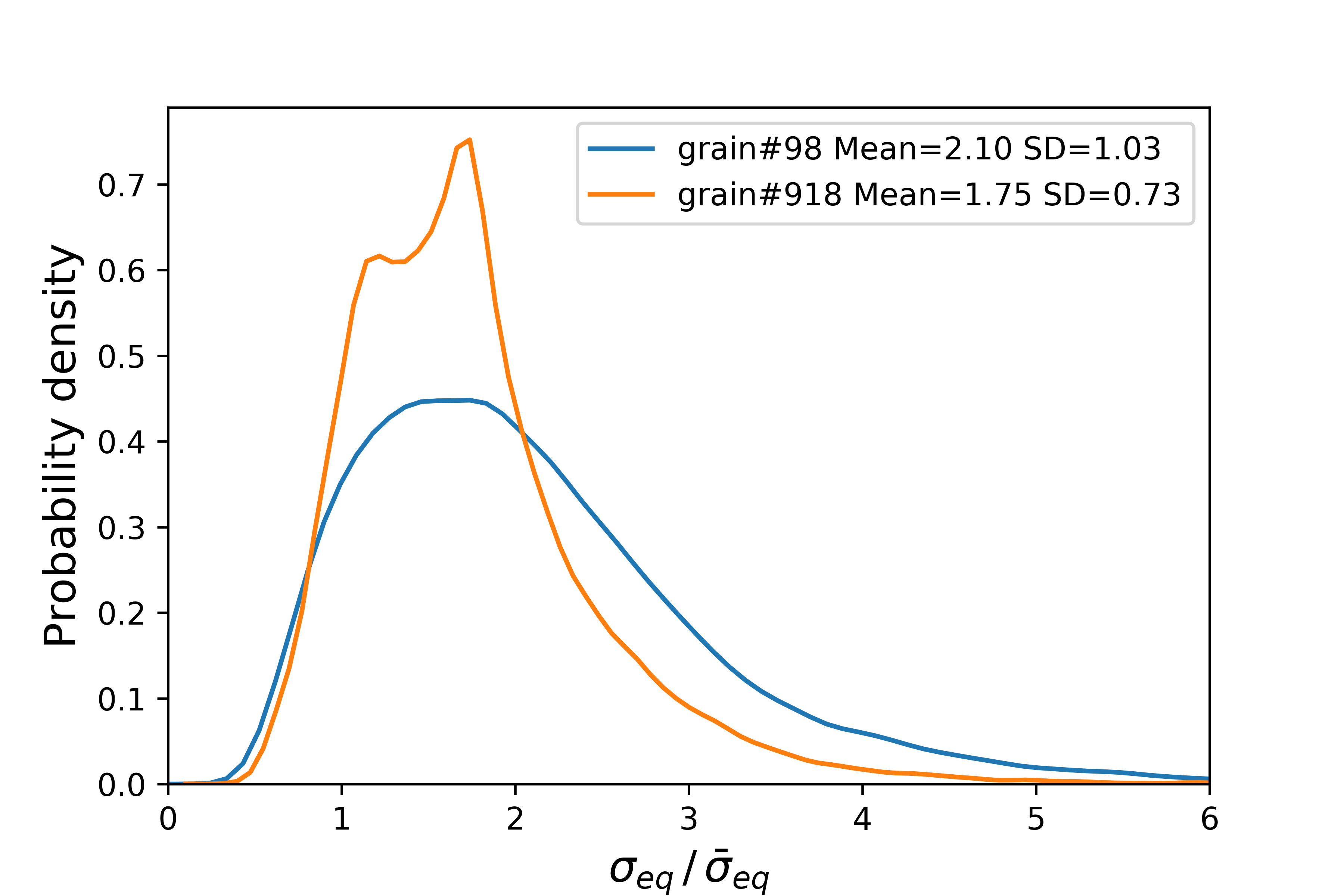}
	\caption{Distribution of normalized equivalent stress 
		$\sigma_{eq}/{\bar\sigma}_{eq}$ in grains $\#98$ and $\#918$ for (left) one microstructure and (right) average over 20 random microstructures. Mean values and standard deviation are also indicated.}
	\label{fig:wad-2grains}
\end{figure}

\begin{figure}[h!]
	\centering 
	\includegraphics[width=6cm,clip]{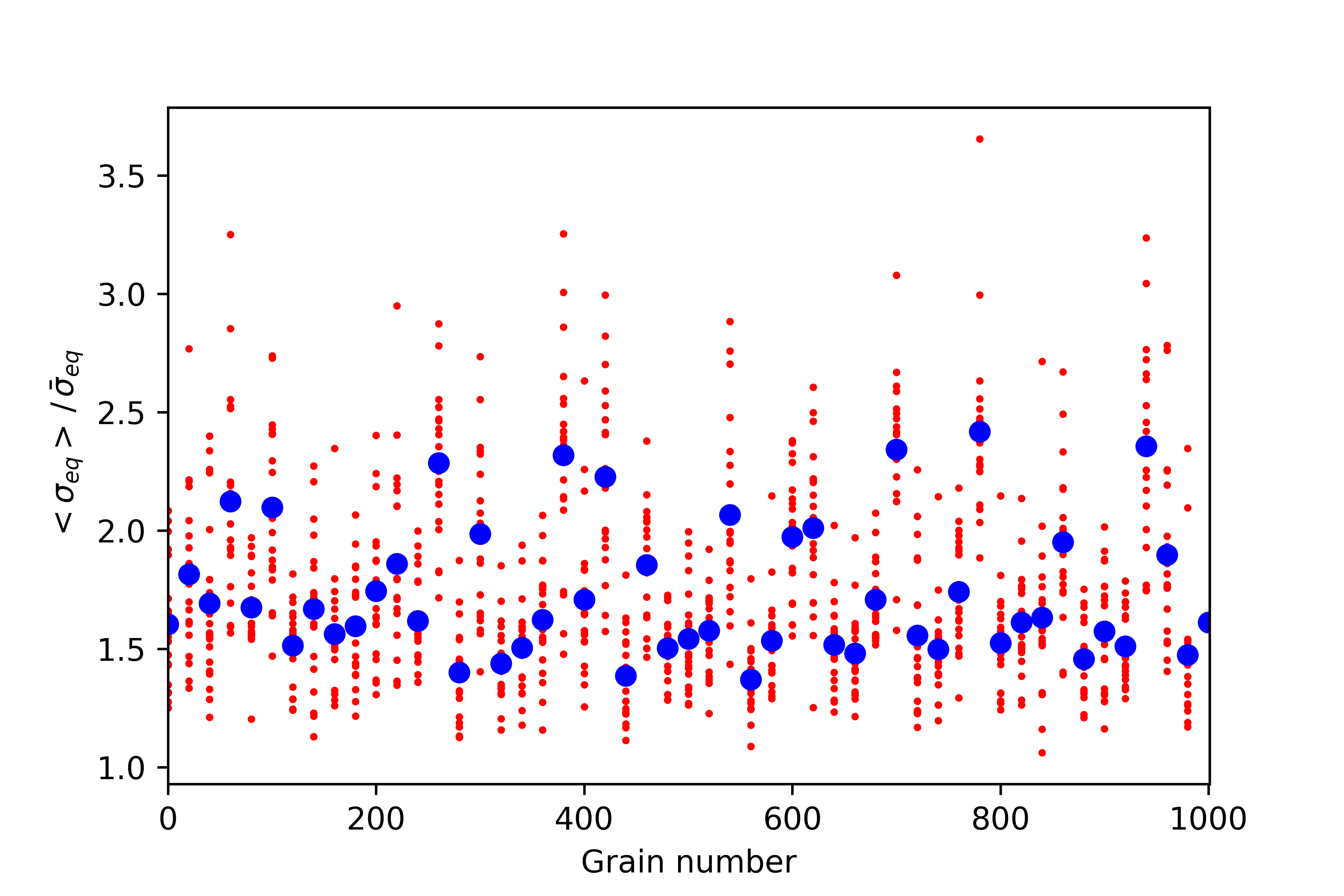}
	\includegraphics[width=6cm,clip]{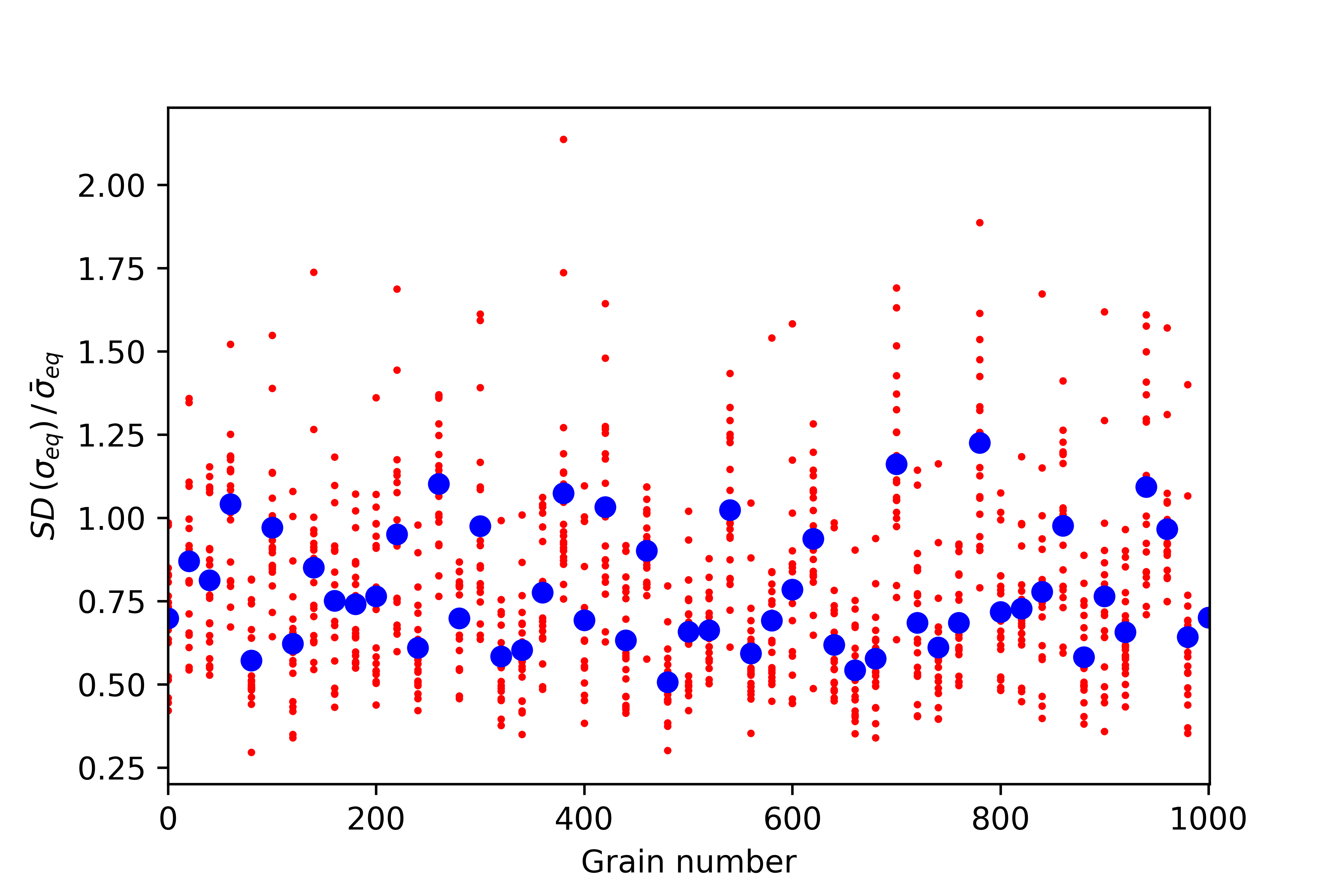}
	\caption{(left) Average values of the local (voxel scale) equivalent stress 
		$<\sigma_{eq}>^{(p)}/{\bar\sigma}_{eq}$ in individual grains, and (right) 
		standard deviation of the equivalent stress 
		$\sqrt{<\sigma_{eq}^2>^{(p)}-(<\sigma_{eq}>^{(p)})^2}/{\bar\sigma}_{eq}$. Only 50 grains are 
		shown out the 1000 grains of the microstructure.}
	\label{fig:wad-moy-SD-grains}
\end{figure}

To be more quantitative, the distribution of equivalent stress for grains $\#98$ and $\#918$ is shown in figure \ref{fig:wad-2grains}.
Figure \ref{fig:wad-2grains}a shows the stress distribution for these grains within a given random microstructure.
One can observe that they are very different, with an intense narrow peak for grain $\#918$ and a broad flat distribution for grain $\#98$.
None of these can be described by simple functions such as a log-normal.
These are typical features encountered in many grains of the FFT computations. 
Stress heterogeneities in grains are due to the viscoplastic anisotropy of the 
grain of interest, related to its crystal orientation, and the mechanical 
interaction of the grain with its surrounding depending on the behavior of 
neighboring grains, the grain shape, and the overall polycrystal behavior.
To estimate the relative importance of both sources, we have carried out FFT computations on 20 microstructures, for which the grains size, shape, and location was random (based on Voronoi tesselation) but keeping the same set of 1000 crystal orientations.
Doing so, it was possible to compute the stress distribution in grains $\#98$ and $\#918$ for those 20 realizations, in which those grains keep the same orientation but change size, shape, and are embedded in various environments. 
Results are shown in figure \ref{fig:wad-2grains}b.
Stress distributions become more similar in shape but still exhibit different mean and standard deviations.
This analysis was generalized for all grains of the microstructure, and the corresponding mean and standard deviation of $\sigma_{eq}$ are shown in figure  \ref{fig:wad-moy-SD-grains}.
In this figure, the red dots correspond to the value obtained for a given random microstructure and the blue dots are the average over 20 microstructures.
It is observed that the stress distribution in all the grains is very sensitive to the specific microstructure considered.
The stress fluctuation in a given grain due to microstructural effects, indicated by the spread of the red dots, is of similar amplitude than the overall stress fluctuation within the whole polycrystal.
In the highly anisotropic and non-linear wadsleyite, the local mechanical state 
of a grain is thus affected by its crystal orientation but also significantly 
by the behavior of neighboring grains.
A consequence of this is that special care should be taken when interpreting 
experimental observation of grain deformation.
The static bound (STAT in table \ref{tab:various_hnl}), assuming a uniform stress within the whole polycrystal and in which grains deformation can be estimated by the sole knowledge of grains orientation (or associated  Schmid factor), is therefore not adapted.
An experimental illustration of this can be found in 
\cite{Grennerat-et-al-2012} in which, for 2D polycrystalline ice (exhibiting 
only two easy slip systems), no correlation was found between the grain Schmid 
factor and the local deformation.

\begin{figure}[h!]
	\centering \includegraphics[width=6cm,clip]{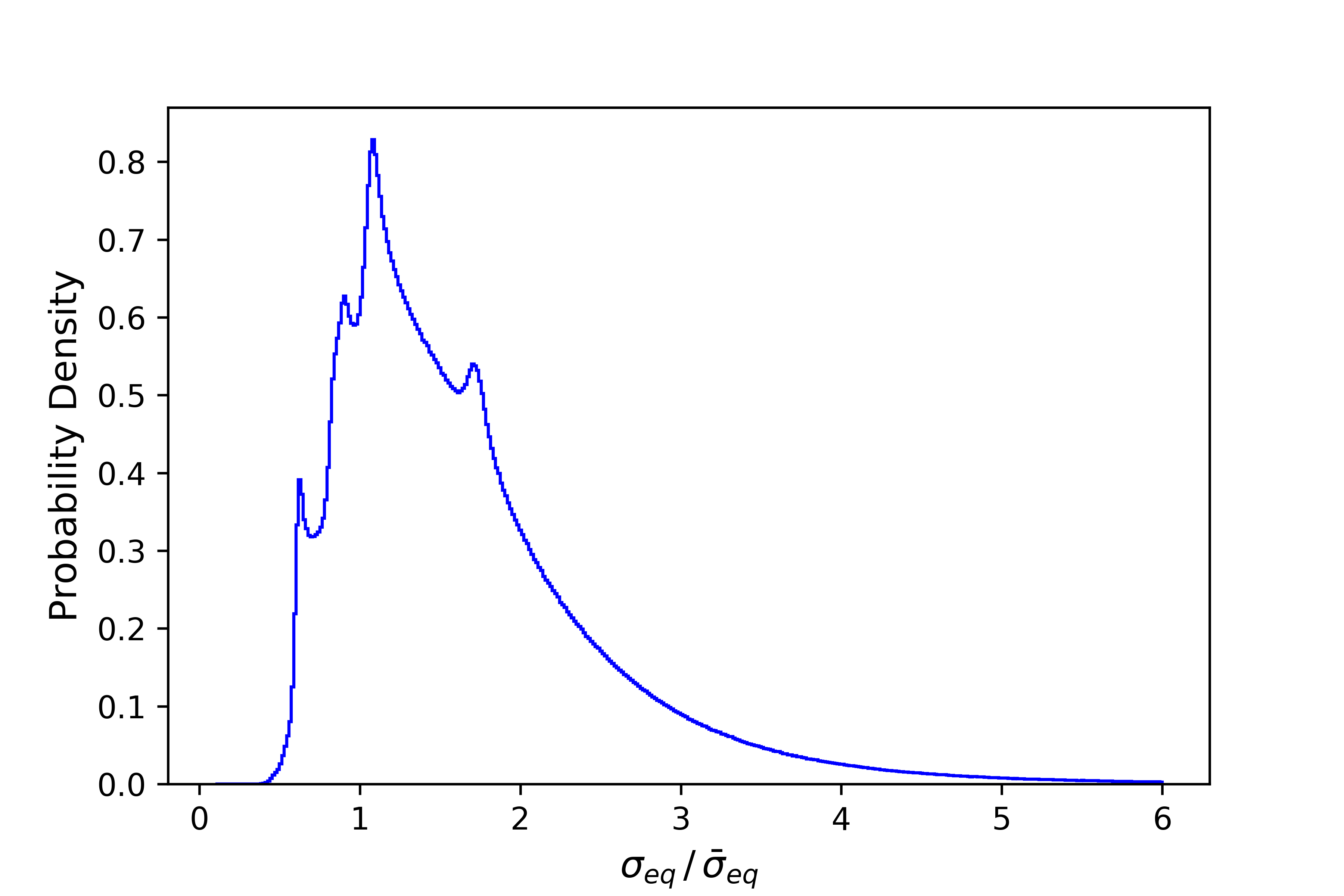}
	\centering \includegraphics[width=6cm,clip]{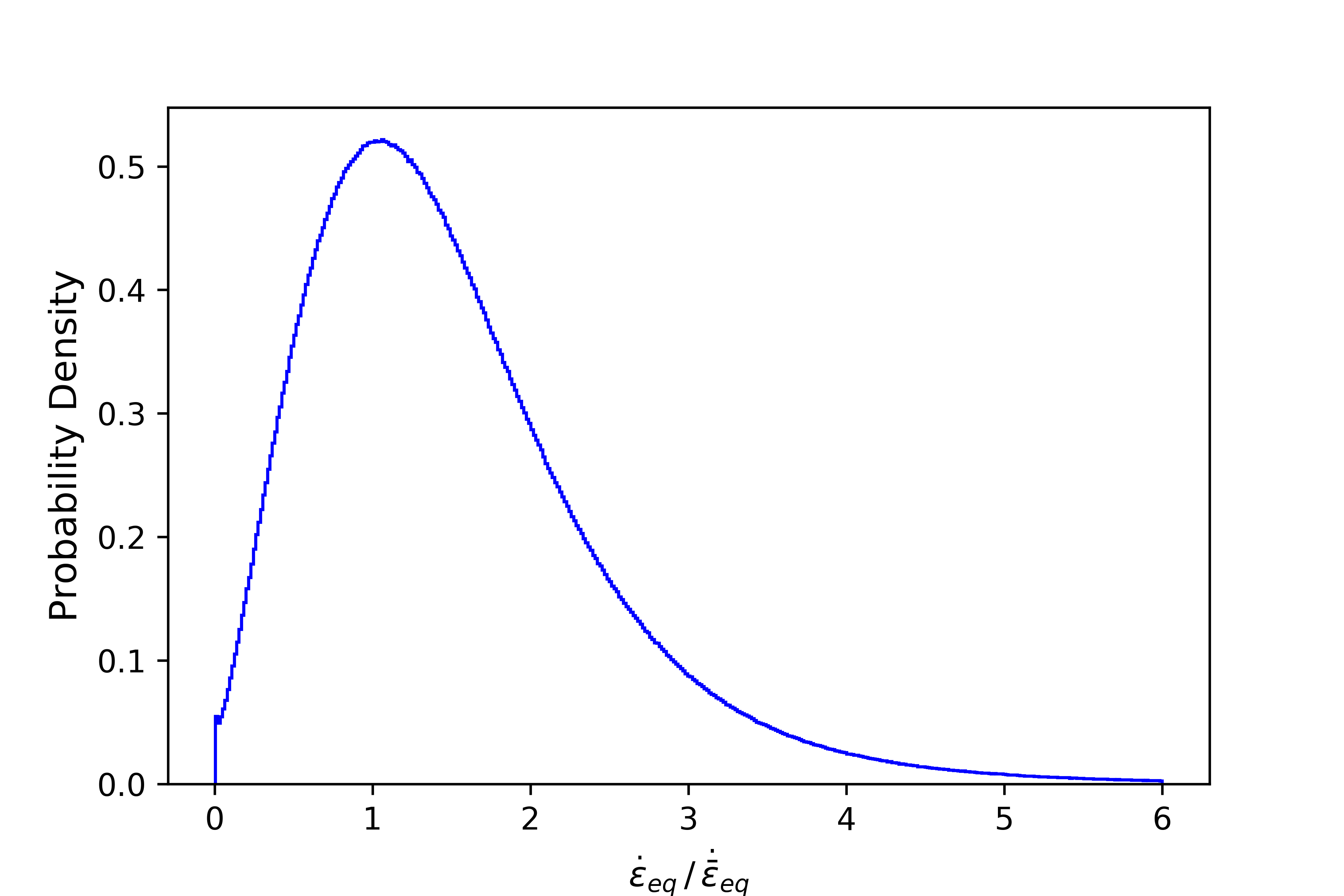}
	\caption{Probability densities of (left) normalized equivalent stress 
		$\sigma_{eq}/{\bar\sigma}_{eq}$ and (right) normalized equivalent 
		strain-rate $\dot\varepsilon_{eq}/\dot{\bar\varepsilon}_{eq}$ predicted by 
		for the whole 3D microstructure (20 random realizations) at $1700\kelvin$.}
	\label{fig:wad-distrib}
\end{figure}

Coming back to the efficiency of FOSO-SC, a challenge for mean-field 
homogenization models is the accurate estimation not only of the effective 
behavior but also of the spread of the mechanical states at the various scales, 
so that good results at the polycrystal scale are obtained for good reasons.
The distribution of normalized stress and strain-rate over the whole polycrystal, 
computed with the FFT homogenization using the 20 random realizations, is given in 
figure \ref{fig:wad-distrib}.
Long tails are observed up to values as high as $\sim 5$.
We have computed the associated standard deviations as these quantities can be also computed by mean-field models.
They are defined as
\begin{equation}
SD(\sigma_{eq}) = \sqrt{ <\sigma_{eq}^2(\mb{x})-\bar\sigma_{eq}^2> } \;, \qquad
SD(\dot\varepsilon_{eq}) = \sqrt{ 
	<\dot\varepsilon_{eq}^2(\mb{x})-\dot{\bar\varepsilon}_{eq}^2> } \;.
\end{equation}
Numerical values are indicated in table \ref{tab:various_hnl}.
It is found that results obtained with FOSO-SC are in very good agrement with the FFT reference ones, the difference being $\sim 10\%$ for the stress heterogeneity and $\sim 15\%$ for strain-rates.
This is an important result as the partially optimized version (POSO) of the SC 
scheme, which was the best available method since 2002, yields results that are 
clearly not as good.
The variational bound (VAR) of \cite{Ponte-Castaneda-1991} predicts stress and 
strain-rate heterogeneities that are here relatively close to the FFT reference 
but, as 
expected, overestimates the effective stress.
Other methods not using the intraphase stress heterogeneities for the 
definition of the linear comparison polycrystals, i.e. the tangent (TGT) 
approach of \cite{Lebensohn-and-Tome-1993} and the affine (AFF) one 
\cite{Masson-et-al-2000} significantly depart from the FFT results, both for 
the effective behavior and the field heterogeneities.
Finally, the static uniform stress bound (STAT) completely discards stress fluctuations and underestimate the flow stress by a factor of $\sim 2$, while the Taylor uniform strain-rate bound (TAYLOR) cannot apply here due to the lack of 5 independent slip systems.

\begin{table}[h!]
	\begin{center}
		\begin{tabular}{|c|c|c|c|c|c|c|c|c|}
			\hline
			&  FFT & STAT & TGT & POSO & FOSO & AFF & VAR & TAYLOR 
			\\
			\hline
			$\bar\sigma_{eq}$  & \textbf{2497} & 1232 & 1699 & 2198 & 2611 & 
			3128 & 3232 
			& $+\infty$ \\  
			$SD(\sigma_{eq})/\bar\sigma_{eq}$       & \textbf{1.59} & 0 & 1.082 
			& 0.856 & 
			1.749 & 2.503 & 1.678 & \\ 
			$SD(\dot\varepsilon_{eq})/\dot{\bar\varepsilon}_{eq}$  &\textbf{ 
				1.495} & 2.898 & 
			1.238 & 2.307 & 1.239 & 2.869 & 1.168 & 0 \\ 
			\hline
		\end{tabular}
		\caption{Effective flow stress $\bar\sigma_{eq}$ and standard deviations (SD) of the 
			equivalent stress and strain-rate as predicted by various extensions 
			of  the SC scheme  at $1700\kelvin$.}
		\label{tab:various_hnl}
	\end{center}
\end{table}

The effective response and associated stress and strain-rate heterogeneities in 
polycrystalline 
aggregates lacking 5 independent slip systems have been studied mostly in HCP 
materials.
Hutchinson \cite{Hutchinson1977} has shown that in the framework of the \textit{linear} SC scheme, used in this study to compute the linear comparison polycrystal, overall polycrystal deformation is possible with only 4 independent systems. 
This result has been shown to apply also for non-linear polycrystals by 
Nebozhyn et al. \cite{Nebozhyn-et-al-2000}.
The same authors also showed that in ionic polycrystals with only 3 independent systems, the variational estimate of \cite{Ponte-Castaneda-1991} predicts a flow stress that is proportional to the square-root of the mechanical contrast between the slip systems, whereas it unrealistically reaches a plateau at high contrasts for the earlier tangent approach of  \cite{Lebensohn-and-Tome-1993}.
A very good match between  the FOSO-SC approach and FFT reference results has 
been obtained in \cite{Song2018} in the case of HCP polycrystals at modest 
non-linearity ($n=3$).
The present study shows that excellent results are also obtained at a 
significantly larger non-linearity.

In Figure \ref{fig:resultats-macro}, our numerical results are compared to the 
available experimental data  in the literature 
\cite{Nishihara2008,Kawazoe2010,Hustoft2013,Farla2015}. 
These experiments take advantage of the latest developments of experimental 
deformation at high $P$, $T$ conditions. Pressure is created by 
compressing the sample assembly between two opposed anvils. This assembly 
contains an internal resistive heater to achieve high temperature conditions. 
Deformation in torsion is produced by rotating one anvil. Stress measurements 
are performed by X-ray diffraction using synchrotron radiation by measuring the 
orientation dependence and changes in lattice spacing for several diffracting 
planes. To ensure acquisition of a proper diffraction pattern, the grain size 
must be maintained small. In all these experiments, the grain size was in the 
range $1-5\;\micro\meter$. 
Although the dispersion of this experimental data set is significant, it 
concludes consistently that wadsleyite is very strong since, despite the high 
temperature involved (ca. $1500-2000\kelvin$), the stress level remains very 
high: 
several GPa. Our modeling results are in excellent agreement with these 
observations. It must be recalled that our multiscale model relies on the 
strong assumption that strain is produced by dislocation glide only. The 
Peierls Nabarro model has shown that lattice friction is strongly affected by 
pressure in this range, partly due to the strong increase of the elastic 
constants under pressure. Despite the small grain size, TEM observations 
\cite{Kawazoe2010,Hustoft2013,Farla2015} agree that dislocation activity is 
pervasive with densities of the order of $10^{13}\;\meter^{-2}$ and dislocation 
configurations which emphasize the key role played by glide under high lattice 
friction (straight dislocations well-confined in glide planes). 
In our 
dislocation based model, glide overcoming high opposed lattice friction leads 
to a strongly non-linear viscoplastic behavior. Experiments provide limited 
constraints on stress 
sensitivity so far, but preliminary estimates from \cite{Hustoft2013,Farla2015} 
suggest stress exponents in the order of $5-6$, i.e. smaller than predicted 
here. Although this evidence needs to 
be consolidated, we can tentatively invoke the role played by grain boundaries. 
As underlined above, the grains sizes in these high-pressure experiments need 
to be maintained in the micrometer range. At the high temperature investigated, 
it is likely that some accommodation processes operate at grain boundaries. 
Indeed, evidence for grain boundary sliding and migration was suggested by 
\cite{Kawazoe2010}. In our polycrystalline models, the grain size is not taken 
into account explicitly, but grain boundaries are present and act only as 
barriers to dislocations glide, leading to strain localizations which cannot 
be accommodated by specific grain boundary relaxation mechanisms. It has been 
demonstrated in \cite{Detrez-et-al-JMPS-2015} that these accommodation 
mechanisms can very efficiently decrease the stress exponent. This might 
reconcile our numerical values to the few experimental data available.

To finish with, we would like to come back to possible geophysical implications for in situ mantle deformation, in which strain-rates are many orders of magnitude smaller than the one considered here. 
Ritterbex et al. \cite{Ritterbex20162085} have provided evidence that dislocation glide controls the mechanical behavior of wadsleyite at lab conditions, and this allows us, in the present paper, to make the link between atomic and grain scales without the need to handle atomic diffusion. 
In contrast, a recent study of Ritterbex et al. \cite{Ritterbex-et-al-2020} handles deformation of wadsleyite at appropriate mantle strain-rates and show that climb-controlled deformation is expected to operate rather than glide-controlled deformation as treated in this work. Therefore, the current results cannot be simply extrapolated to mantle conditions. 
The present work rather aims to provide a theoretical framework which enables to explain the high stress data of deformation experiments of polycrystalline wadsleyite at lab conditions. 
The same procedure could however be applied to these new rheological data, but this is left for future work.


\section{conclusion}

This work provides the very first estimation of the rheology of a 
constituent of the Earth's mantle at relevant pressure and temperature in which 
scales from the atomistic up to polycrystalline aggregate are bridged together 
using state-of-the-art scale transition models.
Here, the dislocation resistance to shear has been computed relying on 
generalized stacking faults energies incorporation the strong influence of 
pressure on atomic bonding, combined with Peierls-Nabarro approach.
The constitutive equation of a slip system involving these dislocations is then 
obtained from the Orowan equation.
It shows that, at high temperature and strain-rates representative of 
laboratory experiments ($10^{-5}\;\second^{-1}$), the rheology at the slip 
system level is highly non-linear due to the high lattice friction opposed to 
dislocation glide.
The obtained constitutive relation at the slip system level has been introduced in the recent Fully Optimized Second Order Self-Consistent scheme (FOSO-SC) of Ponte Casteñeda \cite{PonteCastaneda2015,Song2018} which allows computing the effective viscoplastic behavior at the polycrystal scale. 
Predictions of the latter approach have been compared with the FFT 
computational homogenization method \cite{Moulinec199869} that provides 
reference results, in which the same slip system behavior has been introduced.
We found that the obtained flow stress of wadsleyite polycrystals at 
$15\;\giga\pascal$  matches very well with the available experimental results 
from the literature and lies within experimental uncertainties at least for 
temperatures ranging between $1500\kelvin$ and $2100\kelvin$.
This could be obtained thanks to the fast computation provided by the FOSO-SC method. 
Micromechanical modeling indicates that the stiff slip system $[100](010)$ is 
not activated at all and has no effect on polycrystal deformation.
Wadleyite therefore deforms with only the 4 independent systems of the $1/2<111>\{101\}$ family.
The FFT results show that the deformation of individual 
grains (mean and standard deviation) in the polycrystalline agregate is 
significantly influenced by the behavior of neighboring grains.
Finally, comparison with earlier mean-field approaches demonstrates the superior estimations provided by the FOSO-SC scheme, not only for the effective behavior but also for the overall stress and strain-rate heterogeneities.


\appendix

\section{Determination of the number of independent slip systems}
\label{sec:app1}

Cotton and Kaufman \cite{Cotton1991} proposed a method for determining the number of independent slip systems in cubic crystals.
Here we extend this method to any lattice symmetry.
In that case, the number of independent slip systems not only depends on the Bravais lattice and slip system indexes, but also on the lattice parameters.
For example, out of the six individual slip systems of the family $1/2<110>\{110\}$, only two are independent in the case of a cubic crystal lattice, four are independent in the case of an orthorhombic crystal, but five for a tetragonal crystal.

Consider a crystal lattice with lattice vectors $\mb{a}$, $\mb{b}$ and $\mb{c}$.
The reciprocal lattice vectors $\mb{a}^*$, $\mb{b}^*$ and $\mb{c}^*$ are given by
\begin{equation}
\mb{a}^* = \frac{\mb{b}\times\mb{c}}{V} \;, \qquad
\mb{b}^* = \frac{\mb{c}\times\mb{a}}{V} \;, \qquad
\mb{c}^* = \frac{\mb{a}\times\mb{b}}{V} \;,
\end{equation} 
where $V=(\mb{a},\;\mb{b},\;\mb{c})$ is the lattice volume.
The vector $\mb{n}$ normal to a plane of indexes $(hkl)$ and the vector $\mb{b}$ parallel to the Burgers vector of indexes $[uvw]$ are given by
\begin{equation}
\mb{n} = h\mb{a}^* + k\mb{b}^* + l\mb{c}^* \;, \qquad
\mb{b} = u\mb{a}   + v\mb{b}   + w\mb{c} \;.
\end{equation} 
The strain-rate tensor resulting from dislocations glide on $(hkl)[uvw]$ is 
proportional to the Schmid tensor $\mb{S}$ associated to that system
\begin{equation}
\mb{S}=\frac{1}{2}\left(\hat{\mb{n}}\otimes\hat{\mb{b}}+\hat{\mb{b}}\otimes\hat{\mb{n}}\right)
\end{equation}
where $\hat{\mb{n}}$ and $\hat{\mb{b}}$ denote the 
unit vectors parallel to $\mb{n}$ and $\mb{b}$, respectively.
Tensor $\mb{S}$ is symmetric and traceless ($S_{11}+S_{22}+S_{33}=0$) since plastic deformation due to dislocation glide is isochoric. 
Therefore, it has only five independent components, $S_{11}$, $S_{22}$, $S_{23}$, $S_{13}$ and $S_{12}$.

The number of independent slip systems in the crystal is the rank of the matrix $[M]$ containing as many rows as individual slip systems available in the crystal and five columns filled with the five independent components of $\mb{S}$ for each slip system.
When decomposing $[M]$ in a row echelon form $[M]_r$, for example using the method \texttt{rref} of the python package \texttt{sympy}, the number of independent slip systems corresponds to the number of nonzero rows (rows with at least one nonzero element) of $[M]_r$.

\section{Numerical method for FOSO-SC}
\label{sec:app2}

In this section, we present the numerical method used in 
this work to solve for the FOSO-SC model.
It relies on a linearization step to define the LCC, and an inner loop to solve 
for the LCC.

For the sake of comparison with \cite{Song2018} (their figure 6), we have 
computed 
texture 
development in an ice polycrystal made of 500 grain orientations deformed under 
uniaxial compression up to an overall strain of $150\%$ (in $300$ steps), 
accounting for the evolution of grain shape, and a rheology at the grain level 
with $n=3$ and reference shear stresses $\tau_0$ for prismatic and pyramidal 
slips taken 60 times larger than for basal slip.
Computation with the algorithm below (Fortran90 code) lasts $20 \; \minute$ 
using 
a standard laptop, compared to $20 \;\hour$ as indicated in \cite{Song2018}.
This was obtained using a convergence rate parameter $\kappa=0.5$ (step \ref{algo:5} below) and an 
accuracy $err=10^{-4}$.

For Wadsleyite, the convergence is a little more delicate because of the high 
non-linearity and grain anisotropy. 
We had to use values of $\kappa$ ranging between $0.1$ and $0.005$ while 
increasing the stress sensitivity incrementally (steps of $\Delta n \sim 5$).

\subsection{Iterative linearization of the non-linear polycrystal}
\label{app:nl}
The outer loop of the numerical method works as follows. 
\begin{enumerate}
	\item Initial guess: start computing the uniform stress (static) bound, and use this solution as initial guess for the affine SC model, with the same method as described below but using the appropriated linearization (see section \ref{sec:mean-field}). Then compute the intraphase first $<\bs{\sigma}>^{(p)}$ and second $<\bs{\sigma}\otimes\bs{\sigma}>^{(p)}$ moments of the stress field associated to the affine model (see section \ref{app:lcc}).
	\item \label{algo:2} With the guess values of $<\bs{\sigma}>^{(p)}$ and $<\bs{\sigma}\otimes\bs{\sigma}>^{(p)}$, compute the local compliance $\mb{M}^{(p)}$ and stress-free strain-rate $\bs{\varepsilon}_0^{(p)}$ of the LCC, equation (\ref{eq:local}), according to equations (\ref{eq:tau}), (\ref{eq:loc_constit_rel}) and the FOSO linearization (\ref{eq:lin1}-\ref{eq:tau_hatcheck}).
	\item Solve for the effective behavior ($\tilde{\mb{M}}$, $\tilde{\bs{\varepsilon}}_0$) of the LCC, equation (\ref{eq:effective}), using the method described in section \ref{app:lcc}, and compute the associated new moments $<\bs{\sigma}>^{(p)}_\text{LCC}$ and $<\bs{\sigma}\otimes\bs{\sigma}>^{(p)}_\text{LCC}$.
	\item Invert equation (\ref{eq:effective}) to compute the effective stress $\bar{\bs{\sigma}}$ associated to the prescribed strain-rate $\dot{\bar{\bs{\varepsilon}}}$.
	\item Computed the following four errors:  $\max\limits_{i,j}|<\sigma_{ij}>^{(p)}-\bar\sigma_{ij}|/\max\limits_{k,l}(\bar\sigma_{kl})$, 
	$\max\limits_{i,j}|<\dot\varepsilon_{ij}>^{(p)}-\dot{\bar\varepsilon}_{ij}|/\max\limits_{k,l}(\dot{\bar\varepsilon}_{kl})$,
	$\max\limits_{i,j}|<\sigma_{ij}>^{(p)}_\text{LCC}-<\sigma_{ij}>^{(p)}|/\max\limits_{k,l}(\bar\sigma_{kl})$, and
	$\max\limits_{i,j,k,l}|<\sigma_{ij}\sigma_{kl}>^{(p)}_\text{LCC}-<\sigma_{ij}\sigma_{kl}>^{(p)}|/\max\limits_{m,n}(\bar\sigma_{mn}^2)$.
	\item \label{algo:5} If the largest of the above errors is larger than 
	$err$, then compute new guesses as 
	$\kappa<\bs{\sigma}>^{(p)}_\text{LCC}+(1-\kappa)<\bs{\sigma}>^{(p)}$ and 
	$\kappa<\bs{\sigma}\otimes 
	\bs{\sigma}>^{(p)}_\text{LCC}+(1-\kappa)<\bs{\sigma}\otimes 
	\bs{\sigma}>^{(p)}$ by mixing the initial guess and the estimation for the 
	LCC, and start again with step \ref{algo:2} above.
	
\end{enumerate}

\subsection{Solving for the thermo-elastic Self-Consistent LCC}
\label{app:lcc}
Following e.g. \cite{Brenner-et-al-2004b} and references therein, for given 
$\mb{M}^{(p)}$ and $\bs{\varepsilon}_0^{(p)}$, the effective compliance of the 
LCC is given by the implicit equation
\begin{equation}
\label{eq:mtilde}
\tilde{\mb{M}} + \mb{M}^* = <(\mb{M}^{(p)} + \mb{M}^*)^{-1}>^{-1}
\end{equation}
which is solved by a standard fixed-point iterative method.
When all grains exhibit the same average shape, the Hill influence tensor $\mb{M}^*$ is defined as
\begin{equation}
\mb{M}^* = \mb{E}:\tilde{\mb{M}} \;, \qquad 
\mb{E} = (\mb{S}_E^{-1}-\mb{I})^{-1}
\end{equation}
with $\mb{I}$ the identity tensor and $\mb{S}_E$ the Eshelby tensor 
\begin{equation}
\mb{S}_E = \mb{P}:\tilde{\mb{M}}^{-1} \;, \quad
\mb{P}=\frac{1}{2\pi|\mb{Z}|} \int^{\pi}_0 \int^{\pi}_0 \mb{H}(\bs{\xi}) \|\mb{Z}^{-1}.\bs{\xi} \|^{-3} \sin \theta \text{d}\theta \text{d}\phi 
\end{equation}
with $H_{ijkl}=\frac{1}{2}(K^{-1}_{ik}\xi_{j}\xi_{l}+K^{-1}_{jk}\xi_{i}\xi_{l})$ related to the acoustic tensor $\mb{K}=\bs{\xi}.\tilde{\mb{M}}^{-1}.\bs{\xi}$ and $\bs{\xi}=(\sin \theta \cos \phi,\sin \theta \sin \phi,\cos \theta)$.
$\mb{Z}$ is a symmetrical second order tensor describing the shape of the ellipsoidal inclusion.
Here, integration of $\mb{P}$ is performed at each iteration of 
(\ref{eq:mtilde}) using a gaussian quadrature method with an increasing number 
of Gauss points (as in \cite{Brenner-et-al-2004b}) until an accuracy of 
$err/10$ is reached.
The implicit equation (\ref{eq:mtilde}) is solved iteratively until a precision of $err/10$.
Once $\tilde{\mb{M}}$ has been solved, one computes the stress concentration 
tensor of the purely elastic problem
\begin{equation}
<\mb{B}>^{(p)} = (\mb{M}^{(p)}+\mb{M}^*)^{-1}:(\tilde{\mb{M}}+\mb{M}^*) \;,
\end{equation}
the effective stress-free strain-rate
\begin{equation}
\dot{\tilde{\bs{\varepsilon}}}_0 = <\dot{\bs{\varepsilon}}_0^{(p)} :\mb{B}>
\end{equation}
and the phase-average residual stress
\begin{equation}
<\bs{\sigma}_{res}>^{(p)} = (\mb{M}^{(p)}+\mb{M}^*)^{-1}:(\dot{\tilde{\bs{\varepsilon}}}_0-\dot{\bs{\varepsilon}}_0^{(p)}) \;.
\end{equation}
The intraphase first moment of the stress field is
\begin{equation}
<\bs{\sigma}>^{(p)} = <\mb{B}>^{(p)} : \bar{\bs{\sigma}} + 
<\bs{\sigma}_{res}>^{(p)} \;.
\end{equation}
The second moment is given by 
\begin{equation}
<\sigma_{ij}\sigma_{kl}>^{(p)} = \frac{1}{f_p}\left[
(\bar{\bs{\sigma}} \otimes \bar{\bs{\sigma}}) :: \frac{\partial 
	\tilde{\mb{M}}}{\partial M_{ijkl}^{(p)}} +
2 \frac{\partial \dot{\tilde{\bs{\varepsilon}}}_0}{\partial M_{ijkl}^{(p)}} : 
\bar{\bs{\sigma}} +
\left< \dot{\bs{\varepsilon}}_0^{(p)} : \frac{\partial 
	<\bs{\sigma}_{res}>^{(p)}}{\partial M_{ijkl}^{(p)}}
\right>
\right]
\label{eq:last}
\end{equation}
where expressions for the derivatives can be found e.g. in 
\cite{Brenner-et-al-2004b,Lebensohn-et-al-2007,Brenner-et-al-2009}.
Integration of the derivative of the Eshelby tensor entering in 
(\ref{eq:last}) is done using a gaussian quadrature method.


\bibliographystyle{amsplain}
\bibliography{CRAS}

\providecommand{\bysame}{\leavevmode\hbox to3em{\hrulefill}\thinspace}
\providecommand{\MR}{\relax\ifhmode\unskip\space\fi MR }
\providecommand{\MRhref}[2]{%
  \href{http://www.ams.org/mathscinet-getitem?mr=#1}{#2}
}
\providecommand{\href}[2]{#2}
\begin{thebibliography}{10}

\bibitem{Blackman-etal-GJI2017}
D.K. Blackman, D.E. Boyce, O.~Castelnau, P.~R. Dawson, and G.~Laske,
  \emph{Effects of crystal preferred orientation on upper-mantle flow near
  plate boundaries: rheologic feedbacks and seismic anisotropy}, Geophys. J.
  Int. \textbf{210} (2017), no.~3, 1481--1493.

\bibitem{Brenner-et-al-2004b}
R.~Brenner, O.~Castelnau, and L.~Badea, \emph{Mechanical field fluctuations in
  polycrystals estimated by homogenization techniques}, Proc. Roy. Soc. London
  \textbf{A460} (2004), no.~2052, 3589--3612.

\bibitem{Brenner-et-al-2009}
R.~Brenner, R.L. Lebensohn, and O.~Castelnau, \emph{Elastic anisotropy and
  yield surface estimates}, Int. J. Sol. Struct. \textbf{46} (2009),
  3018--3026.

\bibitem{Cailletaud2003}
G.~Cailletaud, S.~Forest, D.~Jeulin, F.~Feyel, I.~Galliet, V.~Mounoury, and
  S.~Quilici, \emph{Some elements of microstructural mechanics}, Comput. Mater.
  Sci. \textbf{27} (2003), 351--374.

\bibitem{PonteCastaneda2015}
P.~Ponte Castañeda, \emph{Fully optimized second-order variational estimates
  for the macroscopic response and field statistics in viscoplastic crystalline
  composites}, Proc. Royal Soc. A \textbf{471} (2015), no.~2184, 20150665.

\bibitem{Castelnau-et-al-2009}
O.~Castelnau, D.~K. Blackman, and T.~W. Becker, \emph{Numerical simulations of
  texture development and associated rheological anisotropy in regions of
  complex mantle flow}, Geophys. Res. Lett. \textbf{36} (2009), no.~L12304.

\bibitem{Castelnau-et-al-2008a}
O.~Castelnau, D.~K. Blackman, R.~A. Lebensohn, and P.~Ponte{\ }Casta{\~n}eda,
  \emph{Micromechanical modelling of the viscoplastic behavior of olivine}, J.
  Geophys. Res. \textbf{113} (2008).

\bibitem{Castelnau-et-al-2010}
O.~Castelnau, P.~Cordier, R.~A. Lebensohn, S.~Merkel, and P.~Raterron,
  \emph{Microstructures and rheology of the earth's upper mantle inferred from
  a multiscale approach}, Comptes Rendus Physique \textbf{11} (2010), 304--315.

\bibitem{Castelnau-et-al-2008b}
O.~Castelnau, R.~A. Lebensohn, P.~Ponte{\ }Casta{\~n}eda, and D.~K. Blackman,
  \emph{Earth mantle rheology inferred from homogenization theories},
  Multi-scale modeling of heterogeneous materials (O.~Cazacu, ed.), John Wiley
  and Sons, 2008, pp.~55--70.

\bibitem{Cotton1991}
J.~D. Cotton and M.~J. Kaufman, \emph{A simplified method for determining the
  number of independent slip systems in crystals}, Scripta Metal. Mater.
  \textbf{25} (1991), 2395--2398.

\bibitem{Das2019}
Shuvrangsu Das and P.~Ponte Castañeda, \emph{A multiphase homogenization model
  for the viscoplastic response of intact sea ice: the effect of porosity and
  crystallographic texture}, J. Multiscale Comput. Engin. \textbf{17} (2019),
  121--150.

\bibitem{Denoual20071915}
C.~Denoual, \emph{Modeling dislocation by coupling peierls-nabarro and element
  free galerkin methods}, Comput. Methods Appl. Mech. Eng. \textbf{96} (2007),
  1915--1923.

\bibitem{Detrez-et-al-JMPS-2015}
F.~Detrez, O.~Castelnau, P.~Cordier, S.~Merkel, and P.~Raterron,
  \emph{Effective viscoplastic behavior of polycrystalline aggregates lacking
  four independent slip systems inferred from homogenization methods;
  application to olivine}, J. Mech. Phys. Solids \textbf{83} (2015), 199--220.

\bibitem{Farla2015}
R.~Farla, G.~Amulele, J.~Girard, N.~Miyajima, and S.-I. Karato,
  \emph{High-pressure and high-temperature deformation experiments on
  polycrystalline wadsleyite using the rotational drickamer apparatus}, Phys.
  Chem. Miner. \textbf{42} (2015), 541–558.

\bibitem{Gilormini-1995b}
P.~Gilormini, \emph{A critical evaluation for various nonlinear extensions of
  the self-consistent model}, Proc. IUTAM Symp. on Micromechanics of Plasticity
  and Damage of Multiphase Materials (S\`{e}vres, France) (A.~Pineau and
  A.~Zaoui, eds.), Kluwer Academic Publishers, 1995, pp.~67--74.

\bibitem{Gilormini-1995a}
\bysame, \emph{Insuffisance de l'extension classique du modèle autocohérent
  au comportement non linéaire}, C. R. Acad. Sci. Paris \textbf{320} (1995),
  no.~Ser. IIb, 115--122.

\bibitem{Grennerat-et-al-2012}
F.~Grennerat, M.~Montagnat, P.~Duval, and O.~Castelnau nd~P.~Vacher,
  \emph{Intragranular strain field in columnar ice during transient creep},
  Acta Mater. \textbf{60} (2012), no.~8, 3655--3666.

\bibitem{Hustoft2013}
J.~Hustoft, G.~Amulele, J.-I. Ando, K.~Otsuka, Z.~Du, Z.~Jin, and S.-I. Karato,
  \emph{Plastic deformation experiments to high strain on mantle transition
  zoneminerals wadsleyite and ringwoodite in the rotational drickamer
  apparatus}, Earth Planet. Sci. Lett. \textbf{361} (2013), 7--15.

\bibitem{Hutchinson1977}
J.~W. Hutchinson, \emph{Creep and plasticity of hexagonal polycrystals as
  related to single crystal slip}, Met. Trans. \textbf{8A} (1977), no.~9,
  1465--1469.

\bibitem{Idiart-and-Ponte-Castaneda-2007a}
M.~Idiart and P.~Ponte{\ }Casta{\~n}eda, \emph{Field statistics in nonlinear
  composites. i. theory}, Proc. Roy. Soc. London \textbf{463} (2007), 183--202.

\bibitem{Idiart-et-al-2006}
M.~I. Idiart, H.~Moulinec, P.~Ponte{\ }Casta{\~n}eda, and P.~Suquet,
  \emph{Macroscopic behavior and field fluctuations in viscoplastic composites:
  Second-order estimates versus full-field simulations}, J. Mech. Phys. Solids
  \textbf{54} (2006), 1029--1063.

\bibitem{Kanit-et-al-2003}
T.~Kanit, S.~Forest, I.~Galliet, V.~Mounoury, and D.~Jeulin,
  \emph{Determination of the size of the representative volume element for
  random composites: statistical and numerical approach}, Int. J. Sol. Struct.
  \textbf{40} (2003), 3647--3679.

\bibitem{Kawazoe2010}
T.~Kawazoe, S.-I. Karato, J.~Ando, Z.~Jing, and K.~Otsukaand~J.W. Hustoft,
  \emph{Shear deformation of polycrystalline wadsleyite up to 2100 k at 14–17
  gpa using a rotational drickamer apparatus (rda)}, J. Geophys. Res.
  \textbf{115} (2010), 1--11.

\bibitem{Koizumi19933483}
H.~Koizumi, H.~O.~K. Kirchner, and T.~Suzuki, \emph{Kink pair nucleation and
  critical shear stress}, Acta Metall. Mater. \textbf{41} (1993), 3483--3493.

\bibitem{Lebensohn-et-al-2011a}
R.~A. Lebensohn, P.~Ponte{\ }Casta{\~n}eda, R.~Brenner, and O.~Castelnau,
  \emph{Full-field vs. homogenization methods to predict
  microstructure-property relations for polycrystalline materials}, Chapter 11
  of Computational Methods for Microstructure-Property Relationships (S.~Ghosh
  and D.~Dimiduk, eds.), Springer, 2011, pp.~393--441.

\bibitem{Lebensohn-and-Tome-1993}
R.~A. Lebensohn and C.~N. Tom{\'{e}}, \emph{A self-consistent anisotropic
  approach for the simulation of plastic deformation and texture development of
  polycrystals: application to zirconium alloys}, Acta Metall. Mater.
  \textbf{41} (1993), no.~9, 2611--2624.

\bibitem{Lebensohn-et-al-2007}
R.~A. Lebensohn, C.~N. Tom\'{e}, and P.~Ponte{\ }Casta{\~n}eda,
  \emph{Self-consistent modeling of the mechanical behavior of viscoplastic
  polycrystals incorporating field fluctuations}, Phil. Mag. \textbf{87}
  (2007), no.~28, 4287--4322.

\bibitem{Liu-et-al-2003b}
Y.~Liu, P.~Gilormini, and P.~Ponte{\ }Castaneda, \emph{Homogenization estimates
  for texture evolution in halite}, Tectonophysics \textbf{406} (2003),
  179--195.

\bibitem{Liu-et-al-2003}
Y.~Liu, P.~Gilormini, and P.~Ponte{\ }Casta{\~n}eda, \emph{Variational
  self-consistent estimates for texture evolution in viscoplastic
  polycrystals}, Acta Mater. \textbf{51} (2003), 5425--5437.

\bibitem{Masson-et-al-2000}
R.~Masson, M.~Bornert, P.~Suquet, and A.~Zaoui, \emph{An affine formulation for
  the prediction of the effective properties of nonlinear composites and
  polycrystals}, J. Mech. Phys. Solids \textbf{48} (2000), 1203--1226.

\bibitem{Metsue20101467}
A.~Metsue, P.~Carrez, C.~Denoual, D.~Mainprice, and P.~Cordier, \emph{Plastic
  deformation of wadsleyite: Iv dislocation core modelling based on the
  peierls-nabarro-galerkin model}, Acta Mater. \textbf{58} (2010), no.~5,
  1467--1478.

\bibitem{Montagner2015}
J.~P. Montagner, \emph{Deep earth structure - upper mantle structure: Global
  isotropic and anisotropic elastic tomography}, Treatise on Geophysics, 2nd
  edition (G.~Schubert, ed.), vol.~1, Oxford-Elsevier, 2015, pp.~613--639.

\bibitem{Moulinec199869}
H.~Moulinec and P.~Suquet, \emph{A numerical method for computing the overall
  response of nonlinear composites with complex microstructure}, Comput.
  Methods Appl. Mech. Eng \textbf{157} (1998), 69--94.

\bibitem{Nebozhyn-et-al-2000}
M.~V. Nebozhyn, P.~Gilormini, and P.~Ponte{\ }Casta{\~n}eda, \emph{Variational
  self-consistent estimates for viscoplastic polycrystals with highly
  anisotropic grains}, Comptes Rendus Mécanique \textbf{328} (2000), no.~Ser.
  IIb, 11--17.

\bibitem{Nishihara2008}
Y.~Nishihara, D.~Tinker, T.~Kawazoe, Y.~Xu, Z.~Jing, K.~N. Matsukage, and S.-I.
  Karato, \emph{Plastic deformation of wadsleyite and olivine at high-pressure
  and high-temperature using a rotational drickamer apparatus (rda)}, Phys.
  Earth Planet. Int. \textbf{170} (2008), no.~3, 156 -- 169, Frontiers and
  Grand Challenges in Mineral Physics of the Deep Mantle.

\bibitem{Ponte2002a737}
P.~Ponte-Castaneda, \emph{Second-order homogenization estimates for nonlinear
  composites incorporating field fluctuations. part 1: Theory}, J. Mech. Phys.
  Solids \textbf{50} (2002), 737--757.

\bibitem{Ponte-Castaneda-1991}
P.~Ponte{\ }Casta{\~n}eda, \emph{The effective mechanical properties of
  nonlinear isotropic composites}, J. Mech. Phys. Solids \textbf{39} (1991),
  45--71.

\bibitem{Ponte-Castaneda-and-Suquet-1998}
P.~Ponte{\ }Casta{\~n}eda and P.~Suquet, \emph{Nonlinear composites}, Adv.
  Appl. Mech. \textbf{34} (1998), 171--302.

\bibitem{Ribe-etal-2019}
N.~M. Ribe, R.~Hielsche, and O.~Castelnau, \emph{An analytical finite-strain
  parametrization for texture evolution in deforming olivine polycrystals},
  Geophys. J. Int. \textbf{216} (2019), 486--514.

\bibitem{Ritterbex20162085}
S.~Ritterbex, P.~Carrez, and P.~Cordier, \emph{Modeling dislocation glide and
  lattice friction in mg2sio4 waldseyite in conditions of the earth's
  transition zone}, Am. Mineralogist \textbf{101} (2016), 2085--2094.

\bibitem{RitterbexPePi2015}
S.~Ritterbex, P.~Carrez, K.~Gouriet, and P.~Cordier, \emph{Modeling dislocation
  glide in mg2sio4 ringwoodite: towards rheology under transition zone
  conditions}, Phys. Earth Planet. Int. \textbf{248} (2015), 20--28.

\bibitem{Ritterbex-et-al-2020}
S.~Ritterbex, Ph. Carrez, and P.~Cordier, \emph{Deformation across the mantle
  transition zone: A theoretical mineral physics view}, Earth Planet. Sci. Let.
  \textbf{547} (2020), 116438.

\bibitem{Ritterbex2018}
S.~Ritterbex, P.~Hirel, and P.~Carrez, \emph{On low temperature glide of
  dissociated $<$ 1 1 0 $>$ dislocations in strontium titanate}, Philosophical
  Magazine \textbf{98} (2018), no.~15, 1397--1411.

\bibitem{Song2018}
D.~Song and P.~Ponte Castañeda, \emph{Fully optimized second-order
  homogenization estimates for the macroscopic response and texture evolution
  of low-symmetry viscoplastic polycrystals}, Int. J. plasticity \textbf{110}
  (2018), 272--293.

\bibitem{Suquet201264}
P.~Suquet, H.~Moulinec, O.~Castelnau, M.~Montagnat, N.~Lahellec, F.~Grennerat,
  P.~Duval, and R.~Brenner, \emph{Multiscale modeling of the mechanical
  behavior of polycrystalline ice under transient creep}, Procedia IUTAM
  \textbf{3} (2012), 64--78.

\bibitem{Tommasi-et-al-2004}
A.~Tommasi, D.~Mainprice, P.~Cordier, C.~Thoraval, and H.~Couvy,
  \emph{Strain-induced seismic anisotropy of wadsleyite polycrystals and flow
  patterns in the mantle transition zone}, J. Geophys. Res.: Solid Earth
  \textbf{109} (2004), no.~B12.

\end{thebibliography}

\end{document}